\documentclass[reprint, superscriptaddress, twocolumn, amsmath, amssymb, aps, prb]{revtex4-2}
\usepackage{graphicx}
\usepackage{dcolumn}
\usepackage{bm}
\usepackage{placeins}
\usepackage{appendix}

\usepackage{upgreek}

\usepackage{amsthm, amsmath, amssymb, lipsum}
\usepackage{mathrsfs}

\usepackage{verbatim}
\usepackage[usenames,dvipsnames]{xcolor}
 \usepackage{amsmath}
 \usepackage{amsfonts}
 \usepackage{amssymb}
 \usepackage[colorlinks=true,citecolor=blue,linkcolor=red]{hyperref}

 \usepackage{bbold}     
 \usepackage[makeroom]{cancel}  
 \usepackage{multirow}    
 \usepackage[normalem]{ulem}        
 \usepackage{array}
 \usepackage{pdfpages}

\usepackage{siunitx}  
\makeatletter
 \AtBeginDocument{\let\LS@rot\@undefined}
 \makeatother

\begin{document}


\title{Presence versus absence of charging energies in PbTe quantum dots}

\author{Yuhao Wang}
\email{equal contribution}
\affiliation{College of Semiconductors, Southern University of Science and Technology, Shenzhen 518055, China}
\affiliation{State Key Laboratory of Low Dimensional Quantum Physics, Department of Physics, Tsinghua University, Beijing 100084, China}

\author{Lining Yang}
\email{equal contribution}
\affiliation{State Key Laboratory of Low Dimensional Quantum Physics, Department of Physics, Tsinghua University, Beijing 100084, China}

\author{Wenyu Song}
\email{equal contribution}
\affiliation{State Key Laboratory of Low Dimensional Quantum Physics, Department of Physics, Tsinghua University, Beijing 100084, China}

\author{Li Chen}
\email{equal contribution}
\affiliation{School of Physics and Optoelectronic Engineering, Guangdong University of Technology, Guangzhou 510006, China}

\author{Zehao Yu}
\affiliation{State Key Laboratory of Low Dimensional Quantum Physics, Department of Physics, Tsinghua University, Beijing 100084, China}

\author{Xinchen He}
\affiliation{State Key Laboratory of Low Dimensional Quantum Physics, Department of Physics, Tsinghua University, Beijing 100084, China}

\author{Zeyu Yan}
\affiliation{State Key Laboratory of Low Dimensional Quantum Physics, Department of Physics, Tsinghua University, Beijing 100084, China}

\author{Jiaye Xu}
\affiliation{State Key Laboratory of Low Dimensional Quantum Physics, Department of Physics, Tsinghua University, Beijing 100084, China}

\author{Ruidong Li}
\affiliation{State Key Laboratory of Low Dimensional Quantum Physics, Department of Physics, Tsinghua University, Beijing 100084, China}

\author{Weizhao Wang}
\affiliation{State Key Laboratory of Low Dimensional Quantum Physics, Department of Physics, Tsinghua University, Beijing 100084, China}

\author{Zonglin Li}
\affiliation{State Key Laboratory of Low Dimensional Quantum Physics, Department of Physics, Tsinghua University, Beijing 100084, China}

\author{Shuai Yang}
\affiliation{State Key Laboratory of Low Dimensional Quantum Physics, Department of Physics, Tsinghua University, Beijing 100084, China}

\author{Shan Zhang}
\affiliation{State Key Laboratory of Low Dimensional Quantum Physics, Department of Physics, Tsinghua University, Beijing 100084, China}

\author{Xiao Feng}
\affiliation{State Key Laboratory of Low Dimensional Quantum Physics, Department of Physics, Tsinghua University, Beijing 100084, China}
\affiliation{Beijing Academy of Quantum Information Sciences, Beijing 100193, China}
\affiliation{Frontier Science Center for Quantum Information, Beijing 100084, China}
\affiliation{Hefei National Laboratory, Hefei 230088, China}

\author{Tiantian Wang}
\affiliation{Beijing Academy of Quantum Information Sciences, Beijing 100193, China}
\affiliation{Hefei National Laboratory, Hefei 230088, China}

\author{Yunyi Zang}
\affiliation{Beijing Academy of Quantum Information Sciences, Beijing 100193, China}
\affiliation{Hefei National Laboratory, Hefei 230088, China}

\author{Lin Li}
\affiliation{Beijing Academy of Quantum Information Sciences, Beijing 100193, China}

\author{Runan Shang}
\affiliation{Beijing Academy of Quantum Information Sciences, Beijing 100193, China}
\affiliation{Hefei National Laboratory, Hefei 230088, China}

\author{Qi-Kun Xue}
\affiliation{State Key Laboratory of Low Dimensional Quantum Physics, Department of Physics, Tsinghua University, Beijing 100084, China}
\affiliation{Beijing Academy of Quantum Information Sciences, Beijing 100193, China}
\affiliation{Frontier Science Center for Quantum Information, Beijing 100084, China}
\affiliation{Hefei National Laboratory, Hefei 230088, China}
\affiliation{Department of Physics, Southern University of Science and Technology, Shenzhen 518055, China}

\author{Ke He}
\email{kehe@tsinghua.edu.cn}
\affiliation{State Key Laboratory of Low Dimensional Quantum Physics, Department of Physics, Tsinghua University, Beijing 100084, China}
\affiliation{Beijing Academy of Quantum Information Sciences, Beijing 100193, China}
\affiliation{Frontier Science Center for Quantum Information, Beijing 100084, China}
\affiliation{Hefei National Laboratory, Hefei 230088, China}

\author{Hao Zhang}
\email{zhanghao@sustech.edu.cn}
\affiliation{College of Semiconductors, Southern University of Science and Technology, Shenzhen 518055, China}


\begin{abstract}

Charging energy ($E_C$) is essential in quantum dot (QD) devices. Previous studies on PbTe QDs have reported both the presence and absence of $E_C$. To resolve this ambiguity, we vary the QD size, i.e. the cross-sectional area of PbTe nanowires, and track the evolution of $E_C$. For large cross-sectional areas ($\sim \SI{16000}{\nm \squared}$), the PbTe QDs exhibit no measurable $E_C$, while quantized levels are well resolved. Decreasing this area successively to 5000, 1500, and $\SI{460}{\nm^2}$, $E_C$ becomes finite and increases to 80, 160, and $\SI{210}{\micro \eV}$, respectively. We further demonstrate the strong tunability of local gates, which can tune the PbTe device from the QD regime to the regime of ballistic transport. These results address concerns regarding the large dielectric constant of PbTe and provide key insights in engineering advanced PbTe quantum devices.

\end{abstract}

\maketitle  

\section{Introduction}

PbTe nanowires are a promising material platform for engineering Majorana zero modes and topological devices \cite{CaoZhanPbTe, Lutchyn2010, Oreg2010}. Owing to its huge dielectric constant, i.e. $\sim$1350 at $\SI{4.2}{\kelvin}$, charge disorder can be effectively screened. Consequently, high-quality ballistic transport has been achieved \cite{Wenyu_QPC, Yuhao_QPC, Wenyu_Disorder, Yuhao_degeneracy, Quantized_Andreev}, confirming its low-disorder nature. Given that disorder has been the major roadblock, its reduction is crucial for Majorana studies \cite{Patrick_Lee_disorder_2012, Prada2012, GoodBadUgly, DasSarma_estimate, DasSarma2021Disorder, Tudor2021Disorder, Loss_Andreev_band}. Despite these advantages, the large dielectric constant also significantly reduces the charging energy ($E_C$) of PbTe quantum dots (QDs). $E_C = e^2/C$ is determined by the dot capacitance $C$ (here, $e$ is the electron charge).  $C$ scales with the dielectric constant of the environment surrounding the dot, i.e. the PbTe barrier regions that connect the dot and the contacts. Some studies have reported the absence of $E_C$ in PbTe QDs \cite{Frolov_PbTe, Zonglin_QD}, while others reveal a finite but small $E_C$ ($\sim 110-\SI{130}{\micro \eV}$) \cite{Fabrizio_PbTe}. For QDs based on III-V nanowires, $E_C$ is on the order of meV. The ambiguity of $E_C$ in PbTe QDs must be resolved since an appropriate and accurate $E_C$ is essential for topological qubits \cite{2016_PRX_milestone, 2017_Box_qubit, 2017_PRB_Scalable}. Another concern regarding the large dielectric constant is its screening of local electrostatic gates. This screening tends to flatten the potential landscape, hindering the QD formation. Local gates in PbTe devices tend to have a global effect, creating challenges for on-demand local potential modulation.

Here, we address these ambiguities raised by the dielectric constant. Regarding the $E_C$ in PbTe QDs, we varied the nanowire cross-sectional area. As the area decreases, the QD becomes smaller in size and $E_C$ increases from zero, i.e. non-detectable, to finite values. The smaller the area, the larger the $E_C$. To demonstrate the local tunability of potential landscape, we improved the gate design to reduce the crosstalk. Consequently, the PbTe nanowire can be tuned from the regime of quantum point contact (QPC) exhibiting ballistic transport to the regime of QD exhibiting resonance peaks. We also performed electrostatic simulations, which verify this tunability. Our results address the concerns regarding the large dielectric constant and can guide the design of tunable PbTe quantum devices such as topological and spin qubits.   

\section{Quantum Dots with and without charging energies}

\begin{figure}[bt]
\includegraphics[width=\columnwidth]{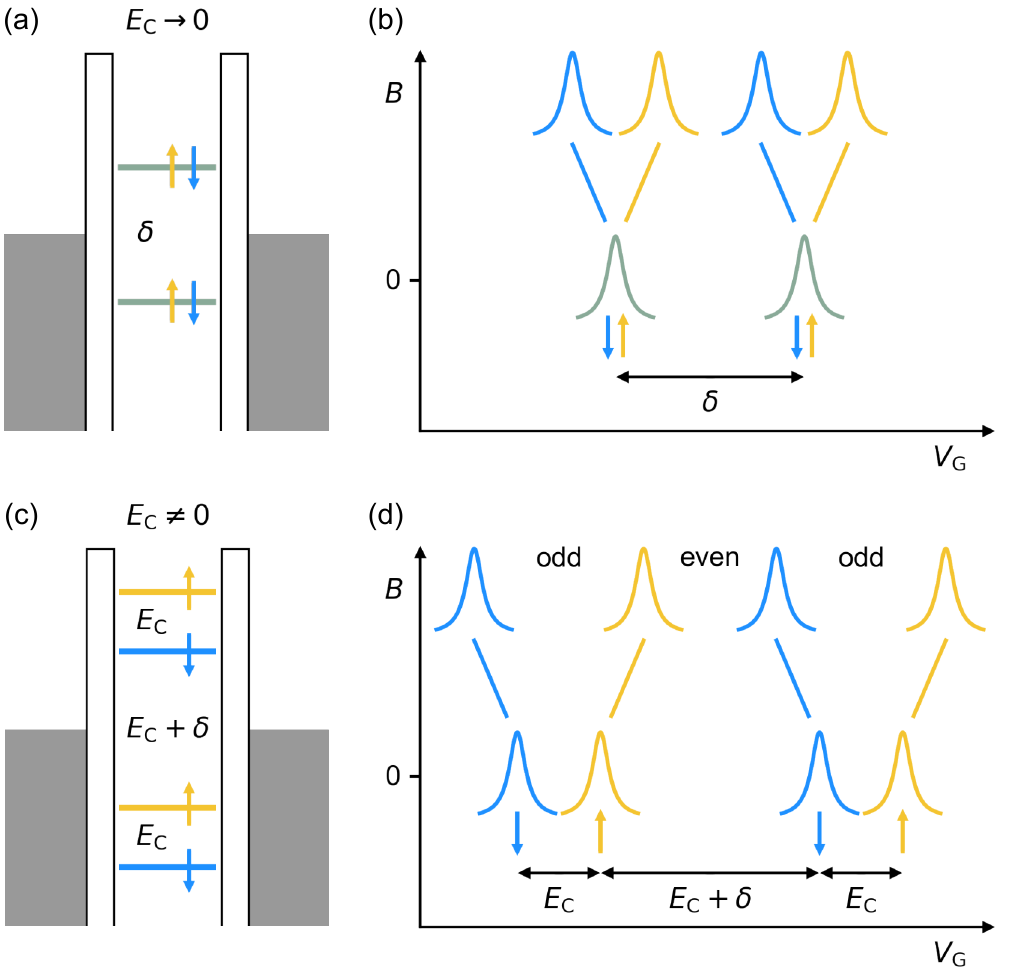}
\centering
\caption{Schematics of QDs with and without $E_C$. (a) Energy diagram of a QD with $E_C$ = 0. $\delta$ denotes the level spacing due to quantum confinement. Arrows indicate spin orientations. (b) Conductance resonances and their evolution with magnetic field. The bias voltage is zero. (c-d) Schematics for the case with $E_C >$ 0.  }
\label{fig1}
\end{figure}

In Figure 1, we sketch the energy diagram of a QD and its transport behavior in the absence ($E_C$ = 0) and presence ($E_C >$ 0) of charging energies. At zero magnetic field ($B$ = 0), the energy spacing between quantized orbital levels is denoted as $\delta$. We assume that the levels do not have orbital degeneracy ($\delta \neq$ 0). When $E_C$ vanishes (Fig. 1(a)), each level is occupied by two electrons with opposite spins simultaneously once it lies below the Fermi energy of the contacts. In transport, a conductance peak is expected to emerge as the gate voltage $V_{\text{G}}$ is swept. A magnetic field induces Zeeman splitting and lifts the spin degeneracy. Consequently, the conductance peak splits as shown in Fig. 1(b).

When $E_C$ is finite, the spin degeneracy at zero field is lifted. An additional energy ($E_C$) is required to add each successive electron, as shown in Fig. 1(c). The energy spacing between consecutive states is either $E_C$ or $E_C + \delta$, corresponding to a Coulomb blockade valley with odd or even number of electrons, respectively. In transport, the spacing between Coulomb peaks alternates between $E_C$ and $E_C+\delta$. As $B$ increases, the odd valley becomes larger and the even valley becomes smaller, see Fig. 1(d) for this evolution.

\section{Charging energies in P\MakeLowercase{b}T\MakeLowercase{e} QD\MakeLowercase{s}}

\begin{figure*}[bt]
\includegraphics[width=\textwidth]{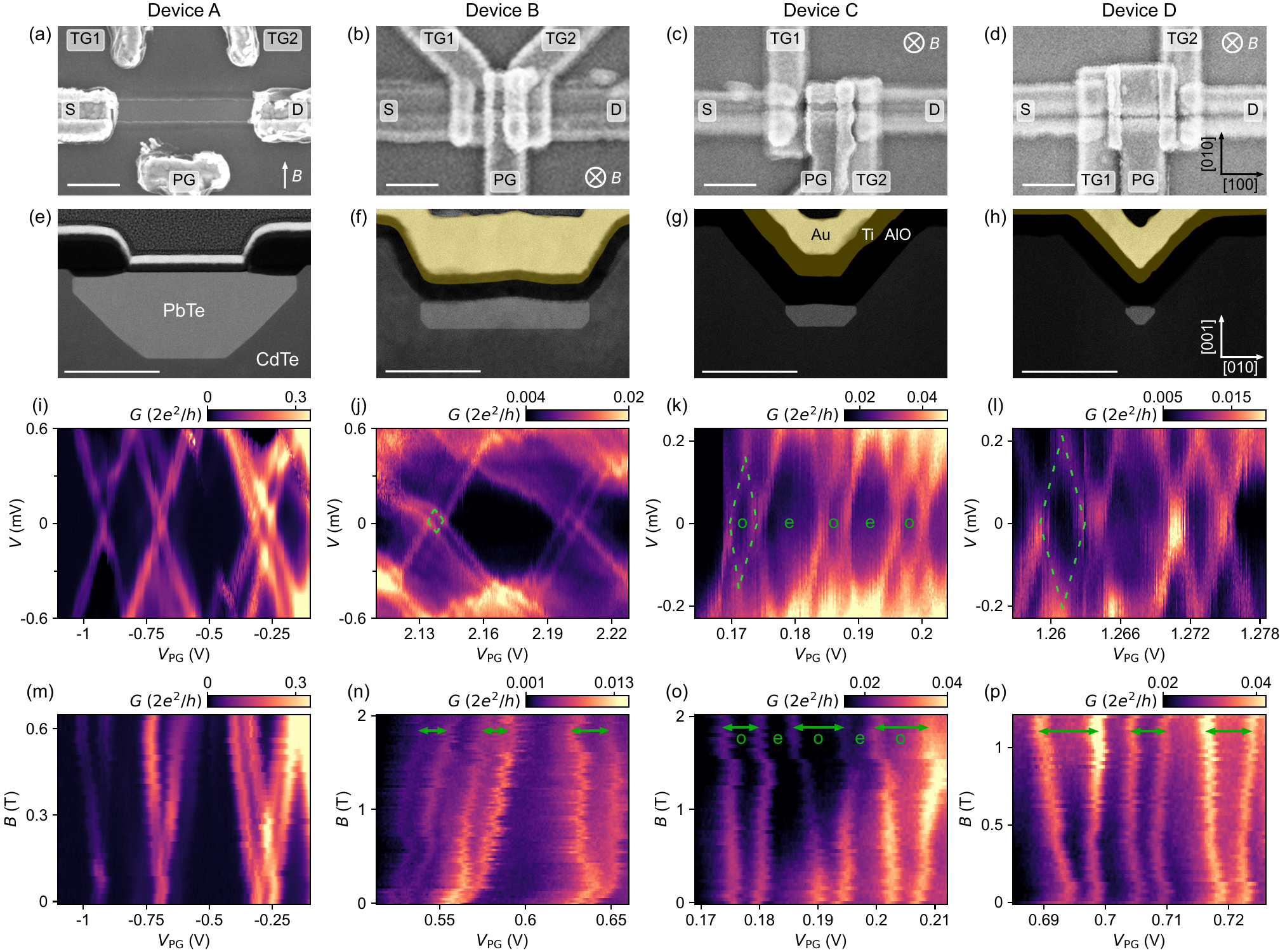}
\centering
\caption{PbTe QD characteristics. (a-d) SEMs of four PbTe QDs. Scale bars are $\SI{300}{\nm}$. (e-h) The corresponding cross-sectional STEMs. Scale bars are $\SI{100}{\nm}$. Device B was burned after its measurement. We therefore show in (f) the STEM of an identical device from the same growth chip. They share the same wire width. The Al$_2$O$_3$ dielectric is labeled as AlO.  (i-l) Charge stability diagrams of the four devices. $B = \SI{0}{\tesla}$. The odd valleys are labeled as "o" and even valleys labeled as "e" in (k). (m-p) $B$ scans of QD states of the four devices. $V = \SI{0}{\mV}$. $B$ directions are labeled in (a-d): Out-of-plane for devices B, C, D and in-plane for device A.  Odd valleys are labeled with green arrows.  }
\label{fig1}
\end{figure*}

\begin{figure}[htb]
\includegraphics[width=\columnwidth]{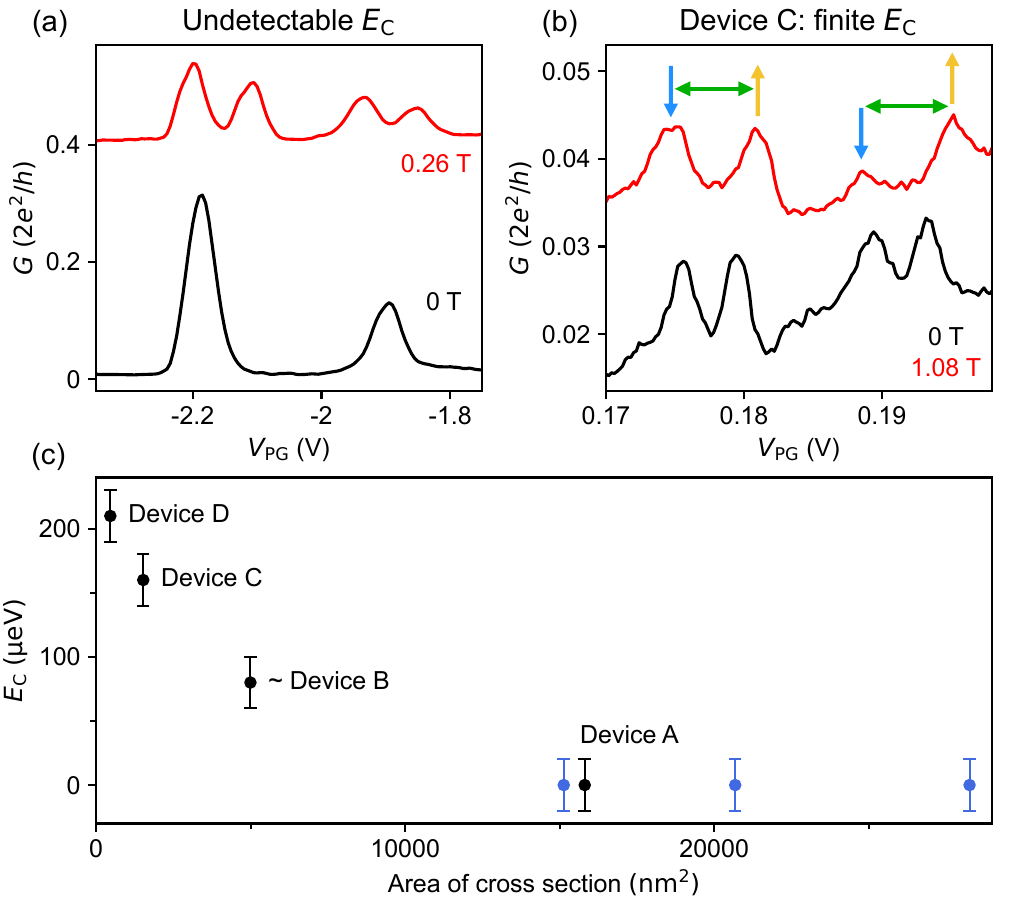}
\centering
\caption{(a) Representative gate scan of a QD device without $E_C$ at $\SI{0}{\tesla}$ (black) and finite $B$ (red). The red curve is offset by 0.4$\times 2e^2/h$ for clarity. (b) Line cuts from Fig. 2(o) showing a finite $E_C$. The red curve is offset by 0.02$\times 2e^2/h$. $V = \SI{0}{\mV}$ for (a-b). (c) Extracted $E_C$ as a function of the cross-sectional area. This area includes the PbEuTe buffer and capping layers. }
\label{fig1}
\end{figure}

We then investigate four PbTe QD devices of different sizes. Figures 2(a-d) show their scanning electron micrographs (SEMs). Detailed characterization of PbTe quantum devices can be found in Refs. \cite{Wenyu_QPC, Yuhao_QPC, Wenyu_Disorder,Yuhao_degeneracy, Quantized_Andreev, Jiangyuying, Erik_PbTe_SAG, PbTe_AB, Zitong, Yichun_Gap, Ruidong_Planar, Vlad_PbTe, PbTe_In, Yichun_SQUID}. Device A has three side gates, while devices B-D implement a top-gate configuration. The two tunnel gates, TG1 and TG2, define the barriers and control the coupling strengths to source and drain contacts. The plunger gate, PG, tunes the electrochemical potential (electron occupancy) inside the QD. To fabricate the top-gate device, a thin layer ($\SIrange{10}{20}{\nm}$ thick) of Al$_2$O$_3$ was deposited via atomic layer deposition followed by the evaporation of TG1 and TG2. A second layer of Al$_2$O$_3$ was deposited, followed by the evaporation of PG. For the SEMs of devices B-D prior to the gate deposition, see the Supplemental Material \cite{SM}.

The cross-sectional area of PbTe was designed to decrease progressively from device A to device D by reducing the trench width (wire width) and PbTe growth time (wire thickness). The scanning transmission electron microscopy (STEM) in Figs. 2(e-h) shows this systematic reduction of cross-sectional areas. From device A to device D, the cross-sectional area is estimated to be $\SI{16000}{\nm^2}$, $\SI{5000}{\nm^2}$, $\SI{1500}{\nm^2}$, and $\SI{460}{\nm^2}$, respectively. The growth and realization of these nanowires will be discussed in a separate work.

Figure 2(i) shows the charge stability diagram of device A, i.e. differential conductance $G\equiv dI/dV$ as a function of $V$ and $V_{\text{PG}}$. Here, $V$ is the bias voltage across the QD, $I$ is the current, $V_{\text{PG}}$ is the gate voltage on PG. The gate voltages on TG1 and TG2, i.e. $V_{\text{TG1}}$ and $V_{\text{TG2}}$, were kept fixed at $\SI{-1.75}{\V}$ and $\SI{-0.55}{\V}$, respectively. In Figure 2(m), we fixed $V$ at $\SI{0}{\mV}$ and scanned $B$. The peaks split as $B$ increases, suggesting the absence of $E_C$ for this device. For the corresponding line cuts, see Fig. S1 \cite{SM}. The diamonds in Fig. 2(i) thus correspond to quantized levels with a level spacing $\delta \sim \SI{600}{\micro \eV}$. We note that "absence of $E_C$" refers to the case where $E_C$ is too small to be resolved within the present measurement resolution. All the measurements were performed in a dilution refrigerator with a base temperature of $\SI{30}{\milli \K}$. Thus, an electron temperature $T_e$ of $\SI{70}{\milli \K}$ corresponds to a thermal broadening of $3.5k_BT_e \sim \SI{20}{\micro \eV}$, where $k_B$ is the Boltzmann constant. We therefore set $\SI{20}{\micro \eV}$ as the measurement uncertainty.

The absence of $E_C$ has been demonstrated in our previous work \cite{Zonglin_QD}. The cross-sectional areas in those devices are either similar or larger than that of device A. From the diamond size, $\SI{0.6}{\mV}$ in $V$ and $\SI{0.2}{\V}$ in $V_{\text{PG}}$, the lever arm of PG can be extracted to be $\sim \SI{3}{\meV \per \V}$. We define the capacitances between the QD and the three gates as $C_{\text{PG}}$, $C_{\text{TG1}}$, and $C_{\text{TG2}}$, and those between the QD and the source and drain contacts as $C_{\text{S}}$ and $C_{\text{D}}$, respectively. Then the total capacitance $C = C_{\text{PG}} + C_{\text{TG1}} +C_{\text{TG2}} +C_{\text{S}} +C_{\text{D}}$ determines the charging energy $E_C = e^2/C$.  The lever arm of PG is $C_{\text{PG}} /C \sim 0.003$, implying that the gate capacitance is orders of magnitude smaller than $C_{\text{S}}$ and $C_{\text{D}}$. Thus, $E_C$ is mainly determined by the contact capacitances. To observe a finite $E_C$, contact capacitance should be reduced. We therefore designed and realized thinner and narrower PbTe nanowires (devices B-D).

Figure 2(f) shows a PbTe nanowire whose cross section is nearly identical to that of device B. The cross-sectional area is roughly one third of that of device A. Consequently, the charge stability diagram (Fig. 2(j)) reveals diamonds with alternating sizes between $\SI{80}{\micro \eV}$ and $\SI{400}{\micro \eV}$. This even-odd pattern is consistent with a finite $E_C$ of $\SI{80}{\micro \eV}$ and $\delta$  of $\SI{320}{\micro \eV}$. We highlight one odd diamond with green dashed lines. To verify this interpretation, Fig. 2(n) shows the $B$ scan of the QD peaks. Unlike the case of device A, the Coulomb peaks do not split in $B$. In addition, the odd valleys (green arrows) grow in peak spacing as $B$ increases while the even valleys shrink, consistent with Figs. 1(c-d). Note that Figs. 2(j) and 2(n) were measured at different gate settings. For their corresponding gate and $B$ scans near the same setting, we refer to Fig. S2 \cite{SM}.

We further decrease the cross-sectional area, see devices C and D in Figs. 2(g-h). The sizes of the odd diamonds in Figs. 2(k) and 2(l), i.e. the charging energies, increase to $\SI{160}{\micro \eV}$ and $\SI{210}{\micro \eV}$, respectively (highlighted by the green dashed diamonds). The $B$ scans in Figs. 2(o-p) show similar behavior with that of device B, further confirming the presence of $E_C$. The frequent fluctuations in $B$ scans are likely due to charge instabilities. We labeled the odd valleys as "o" and even valleys as "e" for device C. The lever arms of PG for devices B-D are extracted to be 0.008, 0.04, and 0.05 (in units of eV/V), respectively. These values are much smaller than 1, which suggests the primary factor affecting $E_C$ still arises from contact capacitances, similar to the side-gate configuration of device A. To achieve larger $E_C$, QDs of smaller cross sections could in principle be grown. Given that the wire width of device D is only $\SI{30}{\nm}$ for the top facet and $\SI{8}{\nm}$ for the bottom facet, wires thinner than this dimension may introduce practical challenges, such as formation of ohmic contacts.  

The size of the odd valley grows in $B$, and this evolution can be used to estimate the g-factor using $\Delta E=g\mu_BB$ ($\mu_B$ is the Bohr magneton). The g-factor for devices B-D are generally below 5 (see Fig. S2 for details), smaller than that of device A ($\sim$8.6). This reduction is probably due to the reduced area and suppressed orbital effect.

To better illustrate the absence versus presence of $E_C$, we compare representative line cuts from two PbTe QDs in Figs. 3(a-b). Figure 3(a) is taken from a device in which $E_C$ is absent, with a cross-sectional area of $\SI{15000}{\nm^2}$. The conductance peaks at zero field (black) split when increasing $B$ (red) due to Zeeman splitting. This behavior is consistent with the schematic shown in Fig. 1(b). For the full $B$ scan of Fig. 3(a), we refer to Fig. S1(b). Figure 3(b) is taken from device C (Fig. 2(o)) with a finite $E_C$. The zero-field peaks do not split. Moreover, the odd valleys with a smaller peak spacing (green arrows) grow in size as $B$ increases, consistent with the schematic shown in Fig. 1(d). For additional data of these QDs, see Figs. S1 and S2  \cite{SM}.

Figure 3(c) summarizes the evolution of $E_C$ as a function of the cross-sectional area. For devices B-D, QD characteristics were measured at two distinct gate settings (shown in Figs. 2 and S2). The value of $E_C$ remains consistent despite variations in gate settings. In addition to the four devices in Fig. 2, three additional devices (shown as blue dots) are included. The transport behavior of these three devices has been investigated in Ref. \cite{Zonglin_QD} with a focus on the g-factor anisotropy. We assign an error bar of $\SI{20}{\micro \eV}$ to account for thermal broadening. The area ranges from $\SI{460}{\nm^2}$ to $\SI{28000}{\nm^2}$, corresponding to a variation of a factor of 60. As the area increases, $E_C$ decreases and eventually becomes non-measurable for areas larger than $\SI{15000}{\nm^2}$. Note that even for QDs with the largest area ($\SI{28000}{\nm^2}$), quantized levels due to confinement can still be well resolved \cite{Zonglin_QD}.  Besides the area, other factors such as the dot length and aspect ratio may also affect the contact capacitance. The aspect ratio of the cross section is governed by the growth dynamics with limited tunability. The dot length (contact spacing) is 270, 390 and $\SI{430}{\nm}$ for devices B-D.  Although this length could, in principle, be further reduced, its tunability is limited by the dielectric thickness and lithographic constraints.

\section{Effect of local gates in P\MakeLowercase{b}T\MakeLowercase{e}}

After resolving the ambiguity of $E_C$, we next investigate the mechanism of QD formation. This is a non-trivial task for PbTe due to its huge dielectric constant and screening effect. A QD can be gate-defined, or formed due to disorder or Schottky barriers at the nanowire-contact interfaces. The former is tunable and desirable, while the latter two are unintentional and poorly controlled. PbTe's screening tends to flatten the potential landscape, hindering gate-defined QDs. This is supported by studies based on PbTe two-dimensional electron gases (2DEGs), where Coulomb blockade or charge quantization is absent in nanostructures with a QD configuration \cite{Physica_E_2004_PbTe_QPC}. Compared with 2DEGs, the nanowire configuration provides stronger confinement. To reduce gate cross talk, we choose top-gate designs with overlapping gates as shown in devices B-D. 

\begin{figure}[htb]
\includegraphics[width=\columnwidth]{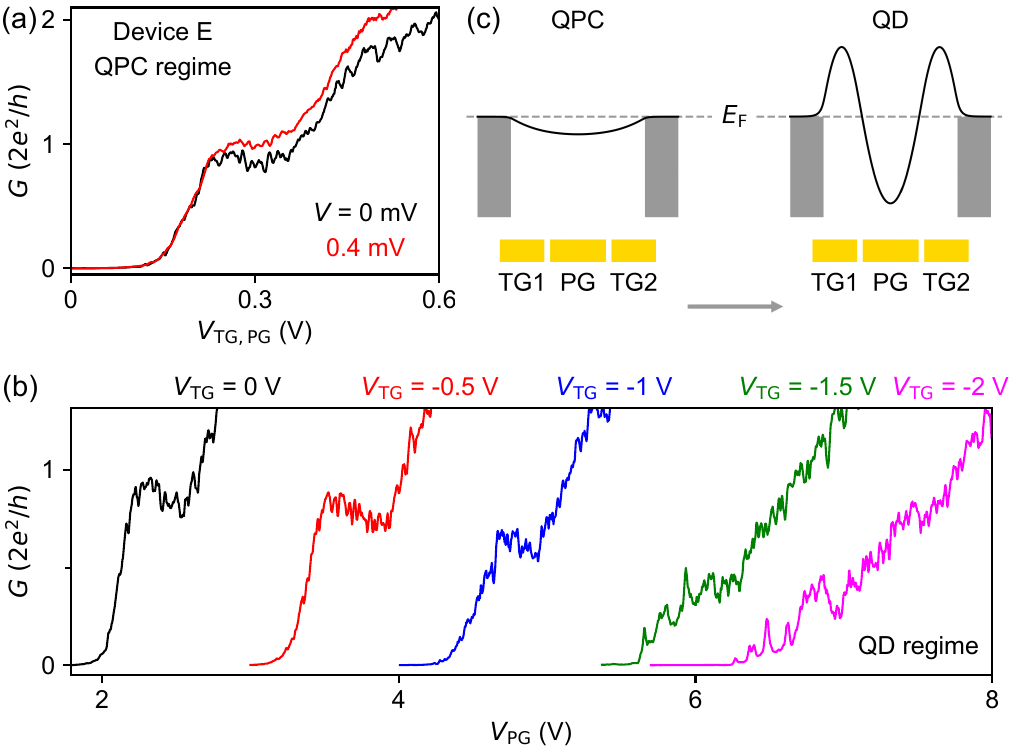}
\centering
\caption{Effects of local gates. (a) Pinch off curves of a QD device. The three gates were swept simultaneously. $B = \SI{0}{\tesla}$. $V = \SI{0.4}{\mV}$ (red) and $\SI{0}{\mV}$ (black). (b) Sweeping $V_{\text{PG}}$ while keeping $V_{\text{TG}}$ fixed at $\SI{0}{\V}$, $\SI{-0.5}{\V}$, $\SI{-1}{\V}$, $\SI{-1.5}{\V}$, and $\SI{-2}{\V}$, respectively. (c) Schematic of the QD potential landscape in the QPC (left) and QD (right) regimes. $V = \SI{0}{\mV}$.}
\label{fig1}
\end{figure}

Figure 4 shows results of such a top-gate QD device (see Fig. S3 for its SEM and STEM). We first scanned the three gates simultaneously ($V_{\text{TG1}}=V_{\text{TG2}}=V_{\text{PG}}$), acting effectively as a single gate like a QPC device. Consequently, conductance quantization, i.e. a plateau near $2e^2/h$, is observed in Fig. 4(a). A contact resistance of 3 k$\Omega$ has been subtracted \cite{Wenyu_Disorder}. The ballistic transport indicates that the potential landscape is nearly flat in this QPC regime, and the disorder level is low and the contacts are transparent. We then fixed $V_{\text{TG1}}=V_{\text{TG2}}=V_{\text{TG}}$ at $\SI{0}{\V}$, and swept $V_{\text{PG}}$. The plateau feature remains (black curve in Fig. 4(b)). Setting $V_{\text{TG}}$ more negative progressively degrades the plateau (see the red and blue curves in Fig. 4(b)). Meanwhile, conductance resonances become more pronounced. For $V_{\text{TG}} = \SI{-2}{\V}$ (the pink curve in Fig. 4(b)), $V_{\text{PG}}$ is near $\SI{+7}{\V}$. This setting drives the device into the QD regime where the potential landscape is no longer flat but has a dip in the middle. Consequently, the ballistic plateau is fully suppressed and Coulomb resonances are ubiquitously observed.

Devices C and D can also exhibit ballistic transport (see Fig. S3), further supporting the gate tunability of local potential landscape. While the disorder level is low, we note that it cannot be fully ruled out and some resonances may be induced by residue disorder or Fabry-P\'{e}rot oscillations from the contacts. The QPC-QD evolution demonstrates that the potential landscape in PbTe is highly tunable using local gates. We illustrate this evolution in Fig. 4(c). As $V_{\text{TG}} = V_{\text{PG}}$, the potential landscape is nearly flat (left panel), and QPC plateau can be expected if the device disorder is low. As $V_{\text{TG}}$ is much more negative than $V_{\text{PG}}$ (right panel), the potential landscape is more QD-like, and the QPC plateau is destroyed. The tunability of local potential landscape with local gates is the key mechanism for gate-induced QDs.

\section{Electrostatic simulations of P\MakeLowercase{b}T\MakeLowercase{e} QD devices}

\begin{figure*}[htb]
\includegraphics[width=0.8\textwidth]{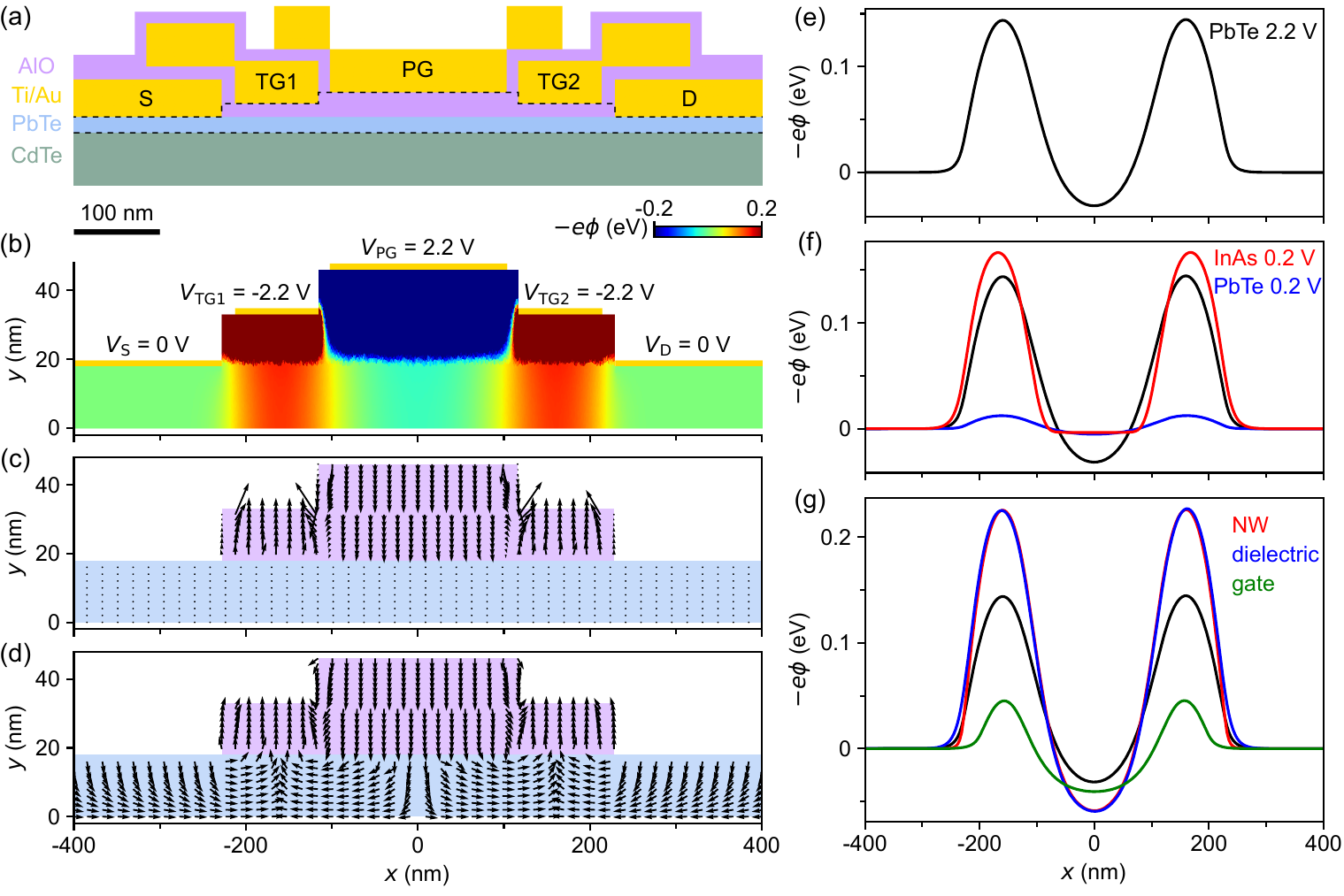}
\centering
\caption{Numerical simulations of the potential landscape. (a) Longitudinal schematic of the a PbTe QD. The scale bar is $\SI{100}{\nm}$. PbTe and PbEuTe are both in light blue.   Source (S) and drain (D) contacts and top gates are in yellow. (b) Potential energy ($-e\phi (x,y)$) of the region in (a) encircled by dashed lines. Note that the $y$-axis is enlarged for clarity. The nanowire segment corresponds to $y < \SI{18}{\nm}$ while the dielectric regions correspond to $y > \SI{18}{\nm}$. The boundary conditions are labeled (yellow lines), $V_{\text{S}}$ = $V_{\text{D}} = \SI{0}{\V}$, $V_{\text{TG1}}$ = $V_{\text{TG2}} = \SI{-2.2}{\V}$, and $V_{\text{PG}} = \SI{2.2}{\V}$. (c) Distribution of the electric field $E(x,y)=-\nabla\phi(x,y)$. The arrow length indicates the field strength while its orientation is the field direction. The field strength in PbTe is so small and the arrows appear as dots. (d) Normalized electric field highlighting its direction. (e) Potential energy distribution at $\phi(x,y=\SI{9}{\nm})$ (the middle of PbTe). (f) The black curve is a replot of (e). The blue curve is the case of setting $V_{\text{TG1}}$ = $V_{\text{TG2}} = \SI{-0.2}{\V}$, $V_{\text{PG}} = \SI{0.2}{\V}$. The red curve is the potential distribution by replacing PbTe with InAs. The gate settings are the same with the blue curve. (g) The black curve is a replot of (e). The blue curve corresponds to halving the thickness of the dielectric layer. The red curve corresponds to halving the PbTe thickness. The green curve corresponds to halving the width of the two tunnel gates. The gate settings remain the same with that in (b).}
\label{fig1}
\end{figure*}

To further elucidate local effects of gates, we show numerical simulations of PbTe QDs in Fig. 5. The QD geometry cut along its longitudinal direction is shown in Fig. 5(a). The PbTe and PbEuTe regions are false-colored blue. For simplicity, we do not differentiate these two regions in our model and treat them both as PbTe. The source and drain contacts and gates set the boundary condition. The model is two-dimensional, i.e. the contacts, PbTe, and gates are assumed to have infinite length along the out-of-page direction.

We calculated the electrostatic potential distribution $\phi(x,y)$ using the two-dimensional Poisson equation,

\begin{equation}
	 \nabla \cdot (\epsilon_0\epsilon_r\nabla\phi(x,y))=-\rho(x,y) 
\label{eq1}
\end{equation}

Here, $\epsilon_0$ is the vacuum dielectric constant, $\epsilon_r(x,y)$ is the relative dielectric constant of the material at position $(x,y)$. $\rho(x,y)$ denotes the mobile charge density which is set to zero in the Al$_2$O$_3$ region. $\rho(x,y)$ in PbTe is obtained using the Thomas-Fermi approximation \cite{LutchynSchrodinger, Flensberg_SP, ChenLi_Simulation}, 

\begin{equation}
\begin{aligned}
	\rho(x,y)=-\frac{4|e|}{3\pi^2\hbar^3} [-2m^e_dE_c(x,y)\Theta(-E_c(x,y))]^{3/2}+ \\ \frac{4|e|}{3\pi^2\hbar^3}[2m^h_dE_v(x,y)\Theta(E_v(x,y))]^{3/2}
\end{aligned}
\label{eq2}
\end{equation}

Here, $m^{e(h)}_d = (m^{e(h)}_l)^{1/3}(m^{e(h)}_t)^{2/3}$ is the density-of-state effective mass for semiconductors with ellipsoidal energy surfaces, where $m^{e(h)}_l$ and $m^{e(h)}_t$ are the longitudinal and transverse effective masses of the electrons (holes) in PbTe, respectively. $\Theta(x,y)$ denotes the Heaviside step function corresponding to the Fermi-Dirac distribution at zero temperature. The conduction band edge $E_c(x,y)$ in the absence of electrostatic potential is taken as the reference energy, such that $E_c(x,y)=0$ when $\phi(x,y)=0$. In particular, the charge density consists of two parts. The first term in Eq. \ref{eq2} represents the electron density originating from the conduction band and exists only when $E_c(x,y)<0$. The second term represents the hole density and exists only when the valence band edge satisfies $E_v=E_c-E_g >0$, where $E_g$ is the band gap of PbTe. Two Dirichlet boundary conditions, $\phi = V_{\text{TG}}$, and $\phi = V_{\text{PG}}$, are imposed on the tunneling and plunger gate regions in the dielectric layer. In addition, the source and drain contacts at the ends of the nanowire are fixed at zero potential. We solve the Eqs. (\ref{eq1}) and (\ref{eq2}) self-consistently to obtain $\phi(x,y)$. 

The parameters of PbTe are \cite{PbTe_parameter1, PbTe_parameter2}, $\epsilon^{\text{PbTe}}_r=1350$, $E_g = \SI{0.15}{\eV}$, $m^e_l$ = 0.24 $m_e$, $m^e_t$ = 0.024 $m_e$, $m^h_l$ = 0.31 $m_e$, $m^h_t$ = 0.022 $m_e$. For the dielectric layer and InAs, relative dielectric constants are $\epsilon^{\text{AlO}}_r=9$, $\epsilon^{\text{InAs}}_r=15$.

Figure 5(b) shows the result for a representative gate setting: $V_{\text{TG1}}$ = $V_{\text{TG2}} = \SI{-2.2}{\V}$, and $V_{\text{PG}} = \SI{2.2}{\V}$. Majority of the potential drop happens in the Al$_2$O$_3$ layer due to the huge mismatch of dielectric constants. To highlight the potential variation in PbTe, we truncated the color bar to [$\SI{-0.2}{\eV}$, $\SI{0.2}{\eV}$]. For the full-range plot, see Fig. S4 \cite{SM}. The regions beneath TG1 and TG2 becomes red, marking the presence of potential barriers. A longitudinal line cut at the middle of PbTe ($y = \SI{9}{\nm}$) is shown in Fig. 5(e), exhibiting a typical potential landscape for QDs. 

Figure 5(c) plots the electric field distribution $E(x,y)=-\nabla\phi(x,y)$. Owing to screening, the field in PbTe is much weaker than that in Al$_2$O$_3$. We then normalize its amplitude and only focus solely on its direction as shown in Fig. 5(d). The electric field direction is strongly refracted at the PbTe/Al$_2$O$_3$ interface.

For comparison, we replace PbTe in Fig. 5(a) with InAs, and calculate its potential landscape (at $y = \SI{9}{\nm}$), see the red curve in Fig. 5(f). To obtain a similar potential landscape, the gate voltages for the InAs case is scaled by an order of magnitude: $V_{\text{TG1}}=V_{\text{TG2}} = \SI{-0.2}{\V}$, $V_{\text{PG}} = \SI{0.2}{\V}$. If this gate setting was applied to PbTe, the tunnel barriers would be significantly reduced (the blue curve).  This tenfold difference in gate voltages reflects the huge dielectric constant mismatch: 1350 (PbTe) vs 15 (InAs). Nevertheless, the InAs-like QD potential (red) can be reproduced in PbTe (black) using local gates. This addresses the feasibility of gate-defined QDs in PbTe.

In Fig. 5(g) we further investigate other factors affecting the potential landscape in PbTe. The blue curve corresponds to halving the Al$_2$O$_3$ thickness. Consequently, the tunnel barriers are higher and the potential energy inside the dot is lower. Halving the PbTe thickness yields a potential landscape at the center of PbTe  ($y = \SI{4.5}{\nm}$, the red curve) that is nearly identical to the case of dielectric layer being halved. The green curve shows the case if the width of tunnel gates is halved. The tunnel barriers become narrower and sharper. Reducing the gate widths could produce more sharply-defined QDs, which is our future effort for optimization.

\section{Summary}

In summary, we investigated charging energies in PbTe quantum dots. By establishing its dependence to the nanowire cross-sectional area, we address the ambiguity on the presence versus absence of $E_C$. With an improved top-gate design, the PbTe devices can be tuned from the QPC regime to the QD regime, validating the tunability of local potential landscape. The ballistic transport in the QPC regime indicates transparent contacts and a low level of disorder, which holds even for the thinnest nanowire. The tunability of local gates is further corroborated by electrostatic numerical simulations. Although the gate voltages need to be scaled by a factor $10$, QD potential profiles can be created in PbTe. Our results address concerns raised by the large dielectric constant of PbTe, such as the ambiguity of $E_C$ and local tunability of gates. Meanwhile, the merit of this large dielectric constant has been experimentally demonstrated \cite{Wenyu_Disorder, Yuhao_degeneracy, Quantized_Andreev}, i.e. disorder level being significantly lower than that in InAs and InSb nanowires. PbTe thus holds great promise towards the realization of Majorana zero modes and topological qubits \cite{NextSteps, Prada2020, Leo_perspective}.

\section{Data Availability}

The data that support the findings of this article are openly available at \cite{rawdata}.

\section{Acknowledgments}

This work is supported by National Natural Science Foundation of China (92565302) and Quantum Science and Technology-National Science and Technology Major Project (2021ZD0302400). W. S. acknowledges the Postdoctoral Fellowship Program and China Postdoctoral Science Foundation (Grant No. BX20250167 and No. 2025M783399). S.Y. acknowledges the China Postdoctoral Science Foundation (Grant No. 2024M751610) and Postdoctoral Fellowship Program of China Postdoctoral Science Foundation (Grant No. GZC20231368).

\bibliography{mybibfile}

\begin{thebibliography}{41}%
\makeatletter
\providecommand \@ifxundefined [1]{%
 \@ifx{#1\undefined}
}%
\providecommand \@ifnum [1]{%
 \ifnum #1\expandafter \@firstoftwo
 \else \expandafter \@secondoftwo
 \fi
}%
\providecommand \@ifx [1]{%
 \ifx #1\expandafter \@firstoftwo
 \else \expandafter \@secondoftwo
 \fi
}%
\providecommand \natexlab [1]{#1}%
\providecommand \enquote  [1]{``#1''}%
\providecommand \bibnamefont  [1]{#1}%
\providecommand \bibfnamefont [1]{#1}%
\providecommand \citenamefont [1]{#1}%
\providecommand \href@noop [0]{\@secondoftwo}%
\providecommand \href [0]{\begingroup \@sanitize@url \@href}%
\providecommand \@href[1]{\@@startlink{#1}\@@href}%
\providecommand \@@href[1]{\endgroup#1\@@endlink}%
\providecommand \@sanitize@url [0]{\catcode `\\12\catcode `\$12\catcode
  `\&12\catcode `\#12\catcode `\^12\catcode `\_12\catcode `\%12\relax}%
\providecommand \@@startlink[1]{}%
\providecommand \@@endlink[0]{}%
\providecommand \url  [0]{\begingroup\@sanitize@url \@url }%
\providecommand \@url [1]{\endgroup\@href {#1}{\urlprefix }}%
\providecommand \urlprefix  [0]{URL }%
\providecommand \Eprint [0]{\href }%
\providecommand \doibase [0]{https://doi.org/}%
\providecommand \selectlanguage [0]{\@gobble}%
\providecommand \bibinfo  [0]{\@secondoftwo}%
\providecommand \bibfield  [0]{\@secondoftwo}%
\providecommand \translation [1]{[#1]}%
\providecommand \BibitemOpen [0]{}%
\providecommand \bibitemStop [0]{}%
\providecommand \bibitemNoStop [0]{.\EOS\space}%
\providecommand \EOS [0]{\spacefactor3000\relax}%
\providecommand \BibitemShut  [1]{\csname bibitem#1\endcsname}%
\let\auto@bib@innerbib\@empty
\bibitem [{\citenamefont {Cao}\ \emph {et~al.}(2022)\citenamefont {Cao},
  \citenamefont {Liu}, \citenamefont {He}, \citenamefont {Liu}, \citenamefont
  {He},\ and\ \citenamefont {Zhang}}]{CaoZhanPbTe}%
  \BibitemOpen
  \bibfield  {author} {\bibinfo {author} {\bibfnamefont {Z.}~\bibnamefont
  {Cao}}, \bibinfo {author} {\bibfnamefont {D.~E.}\ \bibnamefont {Liu}},
  \bibinfo {author} {\bibfnamefont {W.-X.}\ \bibnamefont {He}}, \bibinfo
  {author} {\bibfnamefont {X.}~\bibnamefont {Liu}}, \bibinfo {author}
  {\bibfnamefont {K.}~\bibnamefont {He}},\ and\ \bibinfo {author}
  {\bibfnamefont {H.}~\bibnamefont {Zhang}},\ }\bibfield  {title} {\bibinfo
  {title} {Numerical study of {P}b{T}e-{P}b hybrid nanowires for engineering
  {M}ajorana zero modes},\ }\href {https://doi.org/10.1103/PhysRevB.105.085424}
  {\bibfield  {journal} {\bibinfo  {journal} {Phys. Rev. B}\ }\textbf {\bibinfo
  {volume} {105}},\ \bibinfo {pages} {085424} (\bibinfo {year}
  {2022})}\BibitemShut {NoStop}%
\bibitem [{\citenamefont {Lutchyn}\ \emph {et~al.}(2010)\citenamefont
  {Lutchyn}, \citenamefont {Sau},\ and\ \citenamefont
  {Das~Sarma}}]{Lutchyn2010}%
  \BibitemOpen
  \bibfield  {author} {\bibinfo {author} {\bibfnamefont {R.~M.}\ \bibnamefont
  {Lutchyn}}, \bibinfo {author} {\bibfnamefont {J.~D.}\ \bibnamefont {Sau}},\
  and\ \bibinfo {author} {\bibfnamefont {S.}~\bibnamefont {Das~Sarma}},\
  }\bibfield  {title} {\bibinfo {title} {Majorana fermions and a topological
  phase transition in semiconductor-superconductor heterostructures},\ }\href
  {https://doi.org/10.1103/PhysRevLett.105.077001} {\bibfield  {journal}
  {\bibinfo  {journal} {Phys. Rev. Lett.}\ }\textbf {\bibinfo {volume} {105}},\
  \bibinfo {pages} {077001} (\bibinfo {year} {2010})}\BibitemShut {NoStop}%
\bibitem [{\citenamefont {Oreg}\ \emph {et~al.}(2010)\citenamefont {Oreg},
  \citenamefont {Refael},\ and\ \citenamefont {von Oppen}}]{Oreg2010}%
  \BibitemOpen
  \bibfield  {author} {\bibinfo {author} {\bibfnamefont {Y.}~\bibnamefont
  {Oreg}}, \bibinfo {author} {\bibfnamefont {G.}~\bibnamefont {Refael}},\ and\
  \bibinfo {author} {\bibfnamefont {F.}~\bibnamefont {von Oppen}},\ }\bibfield
  {title} {\bibinfo {title} {Helical liquids and {M}ajorana bound states in
  quantum wires},\ }\href {https://doi.org/10.1103/PhysRevLett.105.177002}
  {\bibfield  {journal} {\bibinfo  {journal} {Phys. Rev. Lett.}\ }\textbf
  {\bibinfo {volume} {105}},\ \bibinfo {pages} {177002} (\bibinfo {year}
  {2010})}\BibitemShut {NoStop}%
\bibitem [{\citenamefont {Song}\ \emph {et~al.}(2023)\citenamefont {Song},
  \citenamefont {Wang}, \citenamefont {Miao}, \citenamefont {Yu}, \citenamefont
  {Gao}, \citenamefont {Li}, \citenamefont {Yang}, \citenamefont {Chen},
  \citenamefont {Geng}, \citenamefont {Zhang}, \citenamefont {Zhang},
  \citenamefont {Zang}, \citenamefont {Cao}, \citenamefont {Liu}, \citenamefont
  {Shang}, \citenamefont {Feng}, \citenamefont {Li}, \citenamefont {Xue},
  \citenamefont {He},\ and\ \citenamefont {Zhang}}]{Wenyu_QPC}%
  \BibitemOpen
  \bibfield  {author} {\bibinfo {author} {\bibfnamefont {W.}~\bibnamefont
  {Song}}, \bibinfo {author} {\bibfnamefont {Y.}~\bibnamefont {Wang}}, \bibinfo
  {author} {\bibfnamefont {W.}~\bibnamefont {Miao}}, \bibinfo {author}
  {\bibfnamefont {Z.}~\bibnamefont {Yu}}, \bibinfo {author} {\bibfnamefont
  {Y.}~\bibnamefont {Gao}}, \bibinfo {author} {\bibfnamefont {R.}~\bibnamefont
  {Li}}, \bibinfo {author} {\bibfnamefont {S.}~\bibnamefont {Yang}}, \bibinfo
  {author} {\bibfnamefont {F.}~\bibnamefont {Chen}}, \bibinfo {author}
  {\bibfnamefont {Z.}~\bibnamefont {Geng}}, \bibinfo {author} {\bibfnamefont
  {Z.}~\bibnamefont {Zhang}}, \bibinfo {author} {\bibfnamefont
  {S.}~\bibnamefont {Zhang}}, \bibinfo {author} {\bibfnamefont
  {Y.}~\bibnamefont {Zang}}, \bibinfo {author} {\bibfnamefont {Z.}~\bibnamefont
  {Cao}}, \bibinfo {author} {\bibfnamefont {D.~E.}\ \bibnamefont {Liu}},
  \bibinfo {author} {\bibfnamefont {R.}~\bibnamefont {Shang}}, \bibinfo
  {author} {\bibfnamefont {X.}~\bibnamefont {Feng}}, \bibinfo {author}
  {\bibfnamefont {L.}~\bibnamefont {Li}}, \bibinfo {author} {\bibfnamefont
  {Q.-K.}\ \bibnamefont {Xue}}, \bibinfo {author} {\bibfnamefont
  {K.}~\bibnamefont {He}},\ and\ \bibinfo {author} {\bibfnamefont
  {H.}~\bibnamefont {Zhang}},\ }\bibfield  {title} {\bibinfo {title}
  {Conductance quantization in {PbTe} nanowires},\ }\href
  {https://doi.org/10.1103/PhysRevB.108.045426} {\bibfield  {journal} {\bibinfo
   {journal} {Phys. Rev. B}\ }\textbf {\bibinfo {volume} {108}},\ \bibinfo
  {pages} {045426} (\bibinfo {year} {2023})}\BibitemShut {NoStop}%
\bibitem [{\citenamefont {Wang}\ \emph {et~al.}(2023)\citenamefont {Wang},
  \citenamefont {Chen}, \citenamefont {Song}, \citenamefont {Geng},
  \citenamefont {Yu}, \citenamefont {Yang}, \citenamefont {Gao}, \citenamefont
  {Li}, \citenamefont {Yang}, \citenamefont {Miao}, \citenamefont {Xu},
  \citenamefont {Wang}, \citenamefont {Xia}, \citenamefont {Song},
  \citenamefont {Feng}, \citenamefont {Wang}, \citenamefont {Zang},
  \citenamefont {Li}, \citenamefont {Shang}, \citenamefont {Xue}, \citenamefont
  {He},\ and\ \citenamefont {Zhang}}]{Yuhao_QPC}%
  \BibitemOpen
  \bibfield  {author} {\bibinfo {author} {\bibfnamefont {Y.}~\bibnamefont
  {Wang}}, \bibinfo {author} {\bibfnamefont {F.}~\bibnamefont {Chen}}, \bibinfo
  {author} {\bibfnamefont {W.}~\bibnamefont {Song}}, \bibinfo {author}
  {\bibfnamefont {Z.}~\bibnamefont {Geng}}, \bibinfo {author} {\bibfnamefont
  {Z.}~\bibnamefont {Yu}}, \bibinfo {author} {\bibfnamefont {L.}~\bibnamefont
  {Yang}}, \bibinfo {author} {\bibfnamefont {Y.}~\bibnamefont {Gao}}, \bibinfo
  {author} {\bibfnamefont {R.}~\bibnamefont {Li}}, \bibinfo {author}
  {\bibfnamefont {S.}~\bibnamefont {Yang}}, \bibinfo {author} {\bibfnamefont
  {W.}~\bibnamefont {Miao}}, \bibinfo {author} {\bibfnamefont {W.}~\bibnamefont
  {Xu}}, \bibinfo {author} {\bibfnamefont {Z.}~\bibnamefont {Wang}}, \bibinfo
  {author} {\bibfnamefont {Z.}~\bibnamefont {Xia}}, \bibinfo {author}
  {\bibfnamefont {H.-D.}\ \bibnamefont {Song}}, \bibinfo {author}
  {\bibfnamefont {X.}~\bibnamefont {Feng}}, \bibinfo {author} {\bibfnamefont
  {T.}~\bibnamefont {Wang}}, \bibinfo {author} {\bibfnamefont {Y.}~\bibnamefont
  {Zang}}, \bibinfo {author} {\bibfnamefont {L.}~\bibnamefont {Li}}, \bibinfo
  {author} {\bibfnamefont {R.}~\bibnamefont {Shang}}, \bibinfo {author}
  {\bibfnamefont {Q.}~\bibnamefont {Xue}}, \bibinfo {author} {\bibfnamefont
  {K.}~\bibnamefont {He}},\ and\ \bibinfo {author} {\bibfnamefont
  {H.}~\bibnamefont {Zhang}},\ }\bibfield  {title} {\bibinfo {title} {Ballistic
  {PbTe} nanowire devices},\ }\href
  {https://doi.org/10.1021/acs.nanolett.3c03604} {\bibfield  {journal}
  {\bibinfo  {journal} {Nano Letters}\ }\textbf {\bibinfo {volume} {23}},\
  \bibinfo {pages} {11137} (\bibinfo {year} {2023})}\BibitemShut {NoStop}%
\bibitem [{\citenamefont {Song}\ \emph {et~al.}(2025)\citenamefont {Song},
  \citenamefont {Yu}, \citenamefont {Wang}, \citenamefont {Gao}, \citenamefont
  {Li}, \citenamefont {Yang}, \citenamefont {Zhang}, \citenamefont {Geng},
  \citenamefont {Li}, \citenamefont {Wang}, \citenamefont {Chen}, \citenamefont
  {Yang}, \citenamefont {Miao}, \citenamefont {Xu}, \citenamefont {Feng},
  \citenamefont {Wang}, \citenamefont {Zang}, \citenamefont {Li}, \citenamefont
  {Shang}, \citenamefont {Xue}, \citenamefont {He},\ and\ \citenamefont
  {Zhang}}]{Wenyu_Disorder}%
  \BibitemOpen
  \bibfield  {author} {\bibinfo {author} {\bibfnamefont {W.}~\bibnamefont
  {Song}}, \bibinfo {author} {\bibfnamefont {Z.}~\bibnamefont {Yu}}, \bibinfo
  {author} {\bibfnamefont {Y.}~\bibnamefont {Wang}}, \bibinfo {author}
  {\bibfnamefont {Y.}~\bibnamefont {Gao}}, \bibinfo {author} {\bibfnamefont
  {Z.}~\bibnamefont {Li}}, \bibinfo {author} {\bibfnamefont {S.}~\bibnamefont
  {Yang}}, \bibinfo {author} {\bibfnamefont {S.}~\bibnamefont {Zhang}},
  \bibinfo {author} {\bibfnamefont {Z.}~\bibnamefont {Geng}}, \bibinfo {author}
  {\bibfnamefont {R.}~\bibnamefont {Li}}, \bibinfo {author} {\bibfnamefont
  {Z.}~\bibnamefont {Wang}}, \bibinfo {author} {\bibfnamefont {F.}~\bibnamefont
  {Chen}}, \bibinfo {author} {\bibfnamefont {L.}~\bibnamefont {Yang}}, \bibinfo
  {author} {\bibfnamefont {W.}~\bibnamefont {Miao}}, \bibinfo {author}
  {\bibfnamefont {J.}~\bibnamefont {Xu}}, \bibinfo {author} {\bibfnamefont
  {X.}~\bibnamefont {Feng}}, \bibinfo {author} {\bibfnamefont {T.}~\bibnamefont
  {Wang}}, \bibinfo {author} {\bibfnamefont {Y.}~\bibnamefont {Zang}}, \bibinfo
  {author} {\bibfnamefont {L.}~\bibnamefont {Li}}, \bibinfo {author}
  {\bibfnamefont {R.}~\bibnamefont {Shang}}, \bibinfo {author} {\bibfnamefont
  {Q.}~\bibnamefont {Xue}}, \bibinfo {author} {\bibfnamefont {K.}~\bibnamefont
  {He}},\ and\ \bibinfo {author} {\bibfnamefont {H.}~\bibnamefont {Zhang}},\
  }\bibfield  {title} {\bibinfo {title} {Reducing disorder in {PbTe} nanowires
  for {M}ajorana research},\ }\href
  {https://doi.org/10.1021/acs.nanolett.4c05708} {\bibfield  {journal}
  {\bibinfo  {journal} {Nano Letters}\ }\textbf {\bibinfo {volume} {25}},\
  \bibinfo {pages} {2350} (\bibinfo {year} {2025})}\BibitemShut {NoStop}%
\bibitem [{\citenamefont {Wang}\ \emph {et~al.}(2024)\citenamefont {Wang},
  \citenamefont {Song}, \citenamefont {Cao}, \citenamefont {Yu}, \citenamefont
  {Yang}, \citenamefont {Li}, \citenamefont {Gao}, \citenamefont {Li},
  \citenamefont {Chen}, \citenamefont {Geng}, \citenamefont {Yang},
  \citenamefont {Xu}, \citenamefont {Wang}, \citenamefont {Zhang},
  \citenamefont {Feng}, \citenamefont {Wang}, \citenamefont {Zang},
  \citenamefont {Li}, \citenamefont {Shang}, \citenamefont {Xue}, \citenamefont
  {Liu}, \citenamefont {He},\ and\ \citenamefont {Zhang}}]{Yuhao_degeneracy}%
  \BibitemOpen
  \bibfield  {author} {\bibinfo {author} {\bibfnamefont {Y.}~\bibnamefont
  {Wang}}, \bibinfo {author} {\bibfnamefont {W.}~\bibnamefont {Song}}, \bibinfo
  {author} {\bibfnamefont {Z.}~\bibnamefont {Cao}}, \bibinfo {author}
  {\bibfnamefont {Z.}~\bibnamefont {Yu}}, \bibinfo {author} {\bibfnamefont
  {S.}~\bibnamefont {Yang}}, \bibinfo {author} {\bibfnamefont {Z.}~\bibnamefont
  {Li}}, \bibinfo {author} {\bibfnamefont {Y.}~\bibnamefont {Gao}}, \bibinfo
  {author} {\bibfnamefont {R.}~\bibnamefont {Li}}, \bibinfo {author}
  {\bibfnamefont {F.}~\bibnamefont {Chen}}, \bibinfo {author} {\bibfnamefont
  {Z.}~\bibnamefont {Geng}}, \bibinfo {author} {\bibfnamefont {L.}~\bibnamefont
  {Yang}}, \bibinfo {author} {\bibfnamefont {J.}~\bibnamefont {Xu}}, \bibinfo
  {author} {\bibfnamefont {Z.}~\bibnamefont {Wang}}, \bibinfo {author}
  {\bibfnamefont {S.}~\bibnamefont {Zhang}}, \bibinfo {author} {\bibfnamefont
  {X.}~\bibnamefont {Feng}}, \bibinfo {author} {\bibfnamefont {T.}~\bibnamefont
  {Wang}}, \bibinfo {author} {\bibfnamefont {Y.}~\bibnamefont {Zang}}, \bibinfo
  {author} {\bibfnamefont {L.}~\bibnamefont {Li}}, \bibinfo {author}
  {\bibfnamefont {R.}~\bibnamefont {Shang}}, \bibinfo {author} {\bibfnamefont
  {Q.-K.}\ \bibnamefont {Xue}}, \bibinfo {author} {\bibfnamefont {D.~E.}\
  \bibnamefont {Liu}}, \bibinfo {author} {\bibfnamefont {K.}~\bibnamefont
  {He}},\ and\ \bibinfo {author} {\bibfnamefont {H.}~\bibnamefont {Zhang}},\
  }\bibfield  {title} {\bibinfo {title} {Gate-tunable subband degeneracy in
  semiconductor nanowires},\ }\href {https://doi.org/10.1073/pnas.2406884121}
  {\bibfield  {journal} {\bibinfo  {journal} {Proceedings of the National
  Academy of Sciences}\ }\textbf {\bibinfo {volume} {121}},\ \bibinfo {pages}
  {e2406884121} (\bibinfo {year} {2024})}\BibitemShut {NoStop}%
\bibitem [{\citenamefont {Gao}\ \emph {et~al.}(2025)\citenamefont {Gao},
  \citenamefont {Song}, \citenamefont {Wang}, \citenamefont {Geng},
  \citenamefont {Cao}, \citenamefont {Yu}, \citenamefont {Yang}, \citenamefont
  {Xu}, \citenamefont {Chen}, \citenamefont {Li} \emph
  {et~al.}}]{Quantized_Andreev}%
  \BibitemOpen
  \bibfield  {author} {\bibinfo {author} {\bibfnamefont {Y.}~\bibnamefont
  {Gao}}, \bibinfo {author} {\bibfnamefont {W.}~\bibnamefont {Song}}, \bibinfo
  {author} {\bibfnamefont {Y.}~\bibnamefont {Wang}}, \bibinfo {author}
  {\bibfnamefont {Z.}~\bibnamefont {Geng}}, \bibinfo {author} {\bibfnamefont
  {Z.}~\bibnamefont {Cao}}, \bibinfo {author} {\bibfnamefont {Z.}~\bibnamefont
  {Yu}}, \bibinfo {author} {\bibfnamefont {S.}~\bibnamefont {Yang}}, \bibinfo
  {author} {\bibfnamefont {J.}~\bibnamefont {Xu}}, \bibinfo {author}
  {\bibfnamefont {F.}~\bibnamefont {Chen}}, \bibinfo {author} {\bibfnamefont
  {Z.}~\bibnamefont {Li}}, \emph {et~al.},\ }\bibfield  {title} {\bibinfo
  {title} {Quantized {A}ndreev conductance in semiconductor nanowires},\ }\href
  {https://doi.org/10.1103/bwp9-7dsd} {\bibfield  {journal} {\bibinfo
  {journal} {Phys. Rev. Appl.}\ }\textbf {\bibinfo {volume} {23}},\ \bibinfo
  {pages} {L061004} (\bibinfo {year} {2025})}\BibitemShut {NoStop}%
\bibitem [{\citenamefont {Liu}\ \emph {et~al.}(2012)\citenamefont {Liu},
  \citenamefont {Potter}, \citenamefont {Law},\ and\ \citenamefont
  {Lee}}]{Patrick_Lee_disorder_2012}%
  \BibitemOpen
  \bibfield  {author} {\bibinfo {author} {\bibfnamefont {J.}~\bibnamefont
  {Liu}}, \bibinfo {author} {\bibfnamefont {A.~C.}\ \bibnamefont {Potter}},
  \bibinfo {author} {\bibfnamefont {K.~T.}\ \bibnamefont {Law}},\ and\ \bibinfo
  {author} {\bibfnamefont {P.~A.}\ \bibnamefont {Lee}},\ }\bibfield  {title}
  {\bibinfo {title} {Zero-bias peaks in the tunneling conductance of
  spin-orbit-coupled superconducting wires with and without {M}ajorana
  end-states},\ }\href {https://doi.org/10.1103/PhysRevLett.109.267002}
  {\bibfield  {journal} {\bibinfo  {journal} {Phys. Rev. Lett.}\ }\textbf
  {\bibinfo {volume} {109}},\ \bibinfo {pages} {267002} (\bibinfo {year}
  {2012})}\BibitemShut {NoStop}%
\bibitem [{\citenamefont {Prada}\ \emph {et~al.}(2012)\citenamefont {Prada},
  \citenamefont {San-Jose},\ and\ \citenamefont {Aguado}}]{Prada2012}%
  \BibitemOpen
  \bibfield  {author} {\bibinfo {author} {\bibfnamefont {E.}~\bibnamefont
  {Prada}}, \bibinfo {author} {\bibfnamefont {P.}~\bibnamefont {San-Jose}},\
  and\ \bibinfo {author} {\bibfnamefont {R.}~\bibnamefont {Aguado}},\
  }\bibfield  {title} {\bibinfo {title} {Transport spectroscopy of {NS}
  nanowire junctions with {M}ajorana fermions},\ }\href
  {https://doi.org/10.1103/PhysRevB.86.180503} {\bibfield  {journal} {\bibinfo
  {journal} {Physical Review B}\ }\textbf {\bibinfo {volume} {86}},\ \bibinfo
  {pages} {180503} (\bibinfo {year} {2012})}\BibitemShut {NoStop}%
\bibitem [{\citenamefont {Pan}\ and\ \citenamefont
  {Das~Sarma}(2020)}]{GoodBadUgly}%
  \BibitemOpen
  \bibfield  {author} {\bibinfo {author} {\bibfnamefont {H.}~\bibnamefont
  {Pan}}\ and\ \bibinfo {author} {\bibfnamefont {S.}~\bibnamefont
  {Das~Sarma}},\ }\bibfield  {title} {\bibinfo {title} {Physical mechanisms for
  zero-bias conductance peaks in {M}ajorana nanowires},\ }\href
  {https://doi.org/10.1103/PhysRevResearch.2.013377} {\bibfield  {journal}
  {\bibinfo  {journal} {Phys. Rev. Research}\ }\textbf {\bibinfo {volume}
  {2}},\ \bibinfo {pages} {013377} (\bibinfo {year} {2020})}\BibitemShut
  {NoStop}%
\bibitem [{\citenamefont {Ahn}\ \emph {et~al.}(2021)\citenamefont {Ahn},
  \citenamefont {Pan}, \citenamefont {Woods}, \citenamefont {Stanescu},\ and\
  \citenamefont {Das~Sarma}}]{DasSarma_estimate}%
  \BibitemOpen
  \bibfield  {author} {\bibinfo {author} {\bibfnamefont {S.}~\bibnamefont
  {Ahn}}, \bibinfo {author} {\bibfnamefont {H.}~\bibnamefont {Pan}}, \bibinfo
  {author} {\bibfnamefont {B.}~\bibnamefont {Woods}}, \bibinfo {author}
  {\bibfnamefont {T.~D.}\ \bibnamefont {Stanescu}},\ and\ \bibinfo {author}
  {\bibfnamefont {S.}~\bibnamefont {Das~Sarma}},\ }\bibfield  {title} {\bibinfo
  {title} {Estimating disorder and its adverse effects in semiconductor
  majorana nanowires},\ }\href
  {https://doi.org/10.1103/PhysRevMaterials.5.124602} {\bibfield  {journal}
  {\bibinfo  {journal} {Phys. Rev. Materials}\ }\textbf {\bibinfo {volume}
  {5}},\ \bibinfo {pages} {124602} (\bibinfo {year} {2021})}\BibitemShut
  {NoStop}%
\bibitem [{\citenamefont {Das~Sarma}\ and\ \citenamefont
  {Pan}(2021)}]{DasSarma2021Disorder}%
  \BibitemOpen
  \bibfield  {author} {\bibinfo {author} {\bibfnamefont {S.}~\bibnamefont
  {Das~Sarma}}\ and\ \bibinfo {author} {\bibfnamefont {H.}~\bibnamefont
  {Pan}},\ }\bibfield  {title} {\bibinfo {title} {Disorder-induced zero-bias
  peaks in {M}ajorana nanowires},\ }\href
  {https://doi.org/10.1103/PhysRevB.103.195158} {\bibfield  {journal} {\bibinfo
   {journal} {Phys. Rev. B}\ }\textbf {\bibinfo {volume} {103}},\ \bibinfo
  {pages} {195158} (\bibinfo {year} {2021})}\BibitemShut {NoStop}%
\bibitem [{\citenamefont {Zeng}\ \emph {et~al.}(2022)\citenamefont {Zeng},
  \citenamefont {Sharma}, \citenamefont {Tewari},\ and\ \citenamefont
  {Stanescu}}]{Tudor2021Disorder}%
  \BibitemOpen
  \bibfield  {author} {\bibinfo {author} {\bibfnamefont {C.}~\bibnamefont
  {Zeng}}, \bibinfo {author} {\bibfnamefont {G.}~\bibnamefont {Sharma}},
  \bibinfo {author} {\bibfnamefont {S.}~\bibnamefont {Tewari}},\ and\ \bibinfo
  {author} {\bibfnamefont {T.}~\bibnamefont {Stanescu}},\ }\bibfield  {title}
  {\bibinfo {title} {Partially separated {M}ajorana modes in a disordered
  medium},\ }\href {https://doi.org/10.1103/PhysRevB.105.205122} {\bibfield
  {journal} {\bibinfo  {journal} {Phys. Rev. B}\ }\textbf {\bibinfo {volume}
  {105}},\ \bibinfo {pages} {205122} (\bibinfo {year} {2022})}\BibitemShut
  {NoStop}%
\bibitem [{\citenamefont {Hess}\ \emph {et~al.}(2023)\citenamefont {Hess},
  \citenamefont {Legg}, \citenamefont {Loss},\ and\ \citenamefont
  {Klinovaja}}]{Loss_Andreev_band}%
  \BibitemOpen
  \bibfield  {author} {\bibinfo {author} {\bibfnamefont {R.}~\bibnamefont
  {Hess}}, \bibinfo {author} {\bibfnamefont {H.~F.}\ \bibnamefont {Legg}},
  \bibinfo {author} {\bibfnamefont {D.}~\bibnamefont {Loss}},\ and\ \bibinfo
  {author} {\bibfnamefont {J.}~\bibnamefont {Klinovaja}},\ }\bibfield  {title}
  {\bibinfo {title} {Trivial {A}ndreev band mimicking topological bulk gap
  reopening in the nonlocal conductance of long {R}ashba nanowires},\ }\href
  {https://doi.org/10.1103/PhysRevLett.130.207001} {\bibfield  {journal}
  {\bibinfo  {journal} {Phys. Rev. Lett.}\ }\textbf {\bibinfo {volume} {130}},\
  \bibinfo {pages} {207001} (\bibinfo {year} {2023})}\BibitemShut {NoStop}%
\bibitem [{\citenamefont {Gomanko}\ \emph {et~al.}(2022)\citenamefont
  {Gomanko}, \citenamefont {de~Jong}, \citenamefont {Jiang}, \citenamefont
  {Schellingerhout}, \citenamefont {Bakkers},\ and\ \citenamefont
  {Frolov}}]{Frolov_PbTe}%
  \BibitemOpen
  \bibfield  {author} {\bibinfo {author} {\bibfnamefont {M.}~\bibnamefont
  {Gomanko}}, \bibinfo {author} {\bibfnamefont {E.~J.}\ \bibnamefont
  {de~Jong}}, \bibinfo {author} {\bibfnamefont {Y.}~\bibnamefont {Jiang}},
  \bibinfo {author} {\bibfnamefont {S.~G.}\ \bibnamefont {Schellingerhout}},
  \bibinfo {author} {\bibfnamefont {E.~P. A.~M.}\ \bibnamefont {Bakkers}},\
  and\ \bibinfo {author} {\bibfnamefont {S.~M.}\ \bibnamefont {Frolov}},\
  }\bibfield  {title} {\bibinfo {title} {{Spin and Orbital Spectroscopy in the
  Absence of Coulomb Blockade in Lead Telluride Nanowire Quantum Dots}},\
  }\href {https://doi.org/10.21468/SciPostPhys.13.4.089} {\bibfield  {journal}
  {\bibinfo  {journal} {SciPost Phys.}\ }\textbf {\bibinfo {volume} {13}},\
  \bibinfo {pages} {089} (\bibinfo {year} {2022})}\BibitemShut {NoStop}%
\bibitem [{\citenamefont {Li}\ \emph {et~al.}(2025)\citenamefont {Li},
  \citenamefont {Song}, \citenamefont {Zhang}, \citenamefont {Wang},
  \citenamefont {Wang}, \citenamefont {Yu}, \citenamefont {Li}, \citenamefont
  {Yan}, \citenamefont {Xu}, \citenamefont {Gao}, \citenamefont {Yang},
  \citenamefont {Yang}, \citenamefont {Feng}, \citenamefont {Wang},
  \citenamefont {Zang}, \citenamefont {Li}, \citenamefont {Shang},
  \citenamefont {Xue}, \citenamefont {He},\ and\ \citenamefont
  {Zhang}}]{Zonglin_QD}%
  \BibitemOpen
  \bibfield  {author} {\bibinfo {author} {\bibfnamefont {Z.}~\bibnamefont
  {Li}}, \bibinfo {author} {\bibfnamefont {W.}~\bibnamefont {Song}}, \bibinfo
  {author} {\bibfnamefont {S.}~\bibnamefont {Zhang}}, \bibinfo {author}
  {\bibfnamefont {Y.}~\bibnamefont {Wang}}, \bibinfo {author} {\bibfnamefont
  {Z.}~\bibnamefont {Wang}}, \bibinfo {author} {\bibfnamefont {Z.}~\bibnamefont
  {Yu}}, \bibinfo {author} {\bibfnamefont {R.}~\bibnamefont {Li}}, \bibinfo
  {author} {\bibfnamefont {Z.}~\bibnamefont {Yan}}, \bibinfo {author}
  {\bibfnamefont {J.}~\bibnamefont {Xu}}, \bibinfo {author} {\bibfnamefont
  {Y.}~\bibnamefont {Gao}}, \bibinfo {author} {\bibfnamefont {S.}~\bibnamefont
  {Yang}}, \bibinfo {author} {\bibfnamefont {L.}~\bibnamefont {Yang}}, \bibinfo
  {author} {\bibfnamefont {X.}~\bibnamefont {Feng}}, \bibinfo {author}
  {\bibfnamefont {T.}~\bibnamefont {Wang}}, \bibinfo {author} {\bibfnamefont
  {Y.}~\bibnamefont {Zang}}, \bibinfo {author} {\bibfnamefont {L.}~\bibnamefont
  {Li}}, \bibinfo {author} {\bibfnamefont {R.}~\bibnamefont {Shang}}, \bibinfo
  {author} {\bibfnamefont {Q.-K.}\ \bibnamefont {Xue}}, \bibinfo {author}
  {\bibfnamefont {K.}~\bibnamefont {He}},\ and\ \bibinfo {author}
  {\bibfnamefont {H.}~\bibnamefont {Zhang}},\ }\bibfield  {title} {\bibinfo
  {title} {Anisotropy of {PbTe} nanowires with and without a superconductor},\
  }\href {https://doi.org/10.1103/PhysRevB.111.195416} {\bibfield  {journal}
  {\bibinfo  {journal} {Phys. Rev. B}\ }\textbf {\bibinfo {volume} {111}},\
  \bibinfo {pages} {195416} (\bibinfo {year} {2025})}\BibitemShut {NoStop}%
\bibitem [{\citenamefont {ten Kate}\ \emph {et~al.}(2022)\citenamefont {ten
  Kate}, \citenamefont {Ritter}, \citenamefont {Fuhrer}, \citenamefont {Jung},
  \citenamefont {Schellingerhout}, \citenamefont {Bakkers}, \citenamefont
  {Riel},\ and\ \citenamefont {Nichele}}]{Fabrizio_PbTe}%
  \BibitemOpen
  \bibfield  {author} {\bibinfo {author} {\bibfnamefont {S.~C.}\ \bibnamefont
  {ten Kate}}, \bibinfo {author} {\bibfnamefont {M.~F.}\ \bibnamefont
  {Ritter}}, \bibinfo {author} {\bibfnamefont {A.}~\bibnamefont {Fuhrer}},
  \bibinfo {author} {\bibfnamefont {J.}~\bibnamefont {Jung}}, \bibinfo {author}
  {\bibfnamefont {S.~G.}\ \bibnamefont {Schellingerhout}}, \bibinfo {author}
  {\bibfnamefont {E.~P. A.~M.}\ \bibnamefont {Bakkers}}, \bibinfo {author}
  {\bibfnamefont {H.}~\bibnamefont {Riel}},\ and\ \bibinfo {author}
  {\bibfnamefont {F.}~\bibnamefont {Nichele}},\ }\bibfield  {title} {\bibinfo
  {title} {Small charging energies and g-factor anisotropy in {PbTe} quantum
  dots},\ }\href {https://doi.org/10.1021/acs.nanolett.2c01943} {\bibfield
  {journal} {\bibinfo  {journal} {Nano Letters}\ }\textbf {\bibinfo {volume}
  {22}},\ \bibinfo {pages} {7049} (\bibinfo {year} {2022})}\BibitemShut
  {NoStop}%
\bibitem [{\citenamefont {Aasen}\ \emph {et~al.}(2016)\citenamefont {Aasen},
  \citenamefont {Hell}, \citenamefont {Mishmash}, \citenamefont {Higginbotham},
  \citenamefont {Danon}, \citenamefont {Leijnse}, \citenamefont {Jespersen},
  \citenamefont {Folk}, \citenamefont {Marcus}, \citenamefont {Flensberg},\
  and\ \citenamefont {Alicea}}]{2016_PRX_milestone}%
  \BibitemOpen
  \bibfield  {author} {\bibinfo {author} {\bibfnamefont {D.}~\bibnamefont
  {Aasen}}, \bibinfo {author} {\bibfnamefont {M.}~\bibnamefont {Hell}},
  \bibinfo {author} {\bibfnamefont {R.~V.}\ \bibnamefont {Mishmash}}, \bibinfo
  {author} {\bibfnamefont {A.}~\bibnamefont {Higginbotham}}, \bibinfo {author}
  {\bibfnamefont {J.}~\bibnamefont {Danon}}, \bibinfo {author} {\bibfnamefont
  {M.}~\bibnamefont {Leijnse}}, \bibinfo {author} {\bibfnamefont {T.~S.}\
  \bibnamefont {Jespersen}}, \bibinfo {author} {\bibfnamefont {J.~A.}\
  \bibnamefont {Folk}}, \bibinfo {author} {\bibfnamefont {C.~M.}\ \bibnamefont
  {Marcus}}, \bibinfo {author} {\bibfnamefont {K.}~\bibnamefont {Flensberg}},\
  and\ \bibinfo {author} {\bibfnamefont {J.}~\bibnamefont {Alicea}},\
  }\bibfield  {title} {\bibinfo {title} {Milestones toward {M}ajorana-based
  quantum computing},\ }\href {https://doi.org/10.1103/PhysRevX.6.031016}
  {\bibfield  {journal} {\bibinfo  {journal} {Phys. Rev. X}\ }\textbf {\bibinfo
  {volume} {6}},\ \bibinfo {pages} {031016} (\bibinfo {year}
  {2016})}\BibitemShut {NoStop}%
\bibitem [{\citenamefont {Plugge}\ \emph {et~al.}(2017)\citenamefont {Plugge},
  \citenamefont {Rasmussen}, \citenamefont {Egger},\ and\ \citenamefont
  {Flensberg}}]{2017_Box_qubit}%
  \BibitemOpen
  \bibfield  {author} {\bibinfo {author} {\bibfnamefont {S.}~\bibnamefont
  {Plugge}}, \bibinfo {author} {\bibfnamefont {A.}~\bibnamefont {Rasmussen}},
  \bibinfo {author} {\bibfnamefont {R.}~\bibnamefont {Egger}},\ and\ \bibinfo
  {author} {\bibfnamefont {K.}~\bibnamefont {Flensberg}},\ }\bibfield  {title}
  {\bibinfo {title} {Majorana box qubits},\ }\href
  {https://doi.org/10.1088/1367-2630/aa54e1} {\bibfield  {journal} {\bibinfo
  {journal} {New Journal of Physics}\ }\textbf {\bibinfo {volume} {19}},\
  \bibinfo {pages} {012001} (\bibinfo {year} {2017})}\BibitemShut {NoStop}%
\bibitem [{\citenamefont {Karzig}\ \emph {et~al.}(2017)\citenamefont {Karzig},
  \citenamefont {Knapp}, \citenamefont {Lutchyn}, \citenamefont {Bonderson},
  \citenamefont {Hastings}, \citenamefont {Nayak}, \citenamefont {Alicea},
  \citenamefont {Flensberg}, \citenamefont {Plugge}, \citenamefont {Oreg},
  \citenamefont {Marcus},\ and\ \citenamefont {Freedman}}]{2017_PRB_Scalable}%
  \BibitemOpen
  \bibfield  {author} {\bibinfo {author} {\bibfnamefont {T.}~\bibnamefont
  {Karzig}}, \bibinfo {author} {\bibfnamefont {C.}~\bibnamefont {Knapp}},
  \bibinfo {author} {\bibfnamefont {R.~M.}\ \bibnamefont {Lutchyn}}, \bibinfo
  {author} {\bibfnamefont {P.}~\bibnamefont {Bonderson}}, \bibinfo {author}
  {\bibfnamefont {M.~B.}\ \bibnamefont {Hastings}}, \bibinfo {author}
  {\bibfnamefont {C.}~\bibnamefont {Nayak}}, \bibinfo {author} {\bibfnamefont
  {J.}~\bibnamefont {Alicea}}, \bibinfo {author} {\bibfnamefont
  {K.}~\bibnamefont {Flensberg}}, \bibinfo {author} {\bibfnamefont
  {S.}~\bibnamefont {Plugge}}, \bibinfo {author} {\bibfnamefont
  {Y.}~\bibnamefont {Oreg}}, \bibinfo {author} {\bibfnamefont {C.~M.}\
  \bibnamefont {Marcus}},\ and\ \bibinfo {author} {\bibfnamefont {M.~H.}\
  \bibnamefont {Freedman}},\ }\bibfield  {title} {\bibinfo {title} {Scalable
  designs for quasiparticle-poisoning-protected topological quantum computation
  with {M}ajorana zero modes},\ }\href
  {https://doi.org/10.1103/PhysRevB.95.235305} {\bibfield  {journal} {\bibinfo
  {journal} {Phys. Rev. B}\ }\textbf {\bibinfo {volume} {95}},\ \bibinfo
  {pages} {235305} (\bibinfo {year} {2017})}\BibitemShut {NoStop}%
\bibitem [{\citenamefont {Jiang}\ \emph {et~al.}(2022)\citenamefont {Jiang},
  \citenamefont {Yang}, \citenamefont {Li}, \citenamefont {Song}, \citenamefont
  {Miao}, \citenamefont {Tong}, \citenamefont {Geng}, \citenamefont {Gao},
  \citenamefont {Li}, \citenamefont {Chen}, \citenamefont {Zhang},
  \citenamefont {Meng}, \citenamefont {Gu}, \citenamefont {Zhu}, \citenamefont
  {Zang}, \citenamefont {Shang}, \citenamefont {Cao}, \citenamefont {Feng},
  \citenamefont {Xue}, \citenamefont {Liu}, \citenamefont {Zhang},\ and\
  \citenamefont {He}}]{Jiangyuying}%
  \BibitemOpen
  \bibfield  {author} {\bibinfo {author} {\bibfnamefont {Y.}~\bibnamefont
  {Jiang}}, \bibinfo {author} {\bibfnamefont {S.}~\bibnamefont {Yang}},
  \bibinfo {author} {\bibfnamefont {L.}~\bibnamefont {Li}}, \bibinfo {author}
  {\bibfnamefont {W.}~\bibnamefont {Song}}, \bibinfo {author} {\bibfnamefont
  {W.}~\bibnamefont {Miao}}, \bibinfo {author} {\bibfnamefont {B.}~\bibnamefont
  {Tong}}, \bibinfo {author} {\bibfnamefont {Z.}~\bibnamefont {Geng}}, \bibinfo
  {author} {\bibfnamefont {Y.}~\bibnamefont {Gao}}, \bibinfo {author}
  {\bibfnamefont {R.}~\bibnamefont {Li}}, \bibinfo {author} {\bibfnamefont
  {F.}~\bibnamefont {Chen}}, \bibinfo {author} {\bibfnamefont {Q.}~\bibnamefont
  {Zhang}}, \bibinfo {author} {\bibfnamefont {F.}~\bibnamefont {Meng}},
  \bibinfo {author} {\bibfnamefont {L.}~\bibnamefont {Gu}}, \bibinfo {author}
  {\bibfnamefont {K.}~\bibnamefont {Zhu}}, \bibinfo {author} {\bibfnamefont
  {Y.}~\bibnamefont {Zang}}, \bibinfo {author} {\bibfnamefont {R.}~\bibnamefont
  {Shang}}, \bibinfo {author} {\bibfnamefont {Z.}~\bibnamefont {Cao}}, \bibinfo
  {author} {\bibfnamefont {X.}~\bibnamefont {Feng}}, \bibinfo {author}
  {\bibfnamefont {Q.-K.}\ \bibnamefont {Xue}}, \bibinfo {author} {\bibfnamefont
  {D.~E.}\ \bibnamefont {Liu}}, \bibinfo {author} {\bibfnamefont
  {H.}~\bibnamefont {Zhang}},\ and\ \bibinfo {author} {\bibfnamefont
  {K.}~\bibnamefont {He}},\ }\bibfield  {title} {\bibinfo {title} {Selective
  area epitaxy of {P}b{T}e-{P}b hybrid nanowires on a lattice-matched
  substrate},\ }\href {https://doi.org/10.1103/PhysRevMaterials.6.034205}
  {\bibfield  {journal} {\bibinfo  {journal} {Phys. Rev. Materials}\ }\textbf
  {\bibinfo {volume} {6}},\ \bibinfo {pages} {034205} (\bibinfo {year}
  {2022})}\BibitemShut {NoStop}%
\bibitem [{\citenamefont {Jung}\ \emph {et~al.}(2022)\citenamefont {Jung},
  \citenamefont {Schellingerhout}, \citenamefont {Ritter}, \citenamefont {ten
  Kate}, \citenamefont {van~der Molen}, \citenamefont {de~Loijer},
  \citenamefont {Verheijen}, \citenamefont {Riel}, \citenamefont {Nichele},\
  and\ \citenamefont {Bakkers}}]{Erik_PbTe_SAG}%
  \BibitemOpen
  \bibfield  {author} {\bibinfo {author} {\bibfnamefont {J.}~\bibnamefont
  {Jung}}, \bibinfo {author} {\bibfnamefont {S.~G.}\ \bibnamefont
  {Schellingerhout}}, \bibinfo {author} {\bibfnamefont {M.~F.}\ \bibnamefont
  {Ritter}}, \bibinfo {author} {\bibfnamefont {S.~C.}\ \bibnamefont {ten
  Kate}}, \bibinfo {author} {\bibfnamefont {O.~A.}\ \bibnamefont {van~der
  Molen}}, \bibinfo {author} {\bibfnamefont {S.}~\bibnamefont {de~Loijer}},
  \bibinfo {author} {\bibfnamefont {M.~A.}\ \bibnamefont {Verheijen}}, \bibinfo
  {author} {\bibfnamefont {H.}~\bibnamefont {Riel}}, \bibinfo {author}
  {\bibfnamefont {F.}~\bibnamefont {Nichele}},\ and\ \bibinfo {author}
  {\bibfnamefont {E.~P.}\ \bibnamefont {Bakkers}},\ }\bibfield  {title}
  {\bibinfo {title} {Selective area growth of {P}b{T}e nanowire networks on
  {I}n{P}},\ }\href {https://doi.org/https://doi.org/10.1002/adfm.202208974}
  {\bibfield  {journal} {\bibinfo  {journal} {Advanced Functional Materials}\
  }\textbf {\bibinfo {volume} {32}},\ \bibinfo {pages} {2208974} (\bibinfo
  {year} {2022})}\BibitemShut {NoStop}%
\bibitem [{\citenamefont {Geng}\ \emph {et~al.}(2022)\citenamefont {Geng},
  \citenamefont {Zhang}, \citenamefont {Chen}, \citenamefont {Yang},
  \citenamefont {Jiang}, \citenamefont {Gao}, \citenamefont {Tong},
  \citenamefont {Song}, \citenamefont {Miao}, \citenamefont {Li}, \citenamefont
  {Wang}, \citenamefont {Zhang}, \citenamefont {Meng}, \citenamefont {Gu},
  \citenamefont {Zhu}, \citenamefont {Zang}, \citenamefont {Li}, \citenamefont
  {Shang}, \citenamefont {Feng}, \citenamefont {Xue}, \citenamefont {He},\ and\
  \citenamefont {Zhang}}]{PbTe_AB}%
  \BibitemOpen
  \bibfield  {author} {\bibinfo {author} {\bibfnamefont {Z.}~\bibnamefont
  {Geng}}, \bibinfo {author} {\bibfnamefont {Z.}~\bibnamefont {Zhang}},
  \bibinfo {author} {\bibfnamefont {F.}~\bibnamefont {Chen}}, \bibinfo {author}
  {\bibfnamefont {S.}~\bibnamefont {Yang}}, \bibinfo {author} {\bibfnamefont
  {Y.}~\bibnamefont {Jiang}}, \bibinfo {author} {\bibfnamefont
  {Y.}~\bibnamefont {Gao}}, \bibinfo {author} {\bibfnamefont {B.}~\bibnamefont
  {Tong}}, \bibinfo {author} {\bibfnamefont {W.}~\bibnamefont {Song}}, \bibinfo
  {author} {\bibfnamefont {W.}~\bibnamefont {Miao}}, \bibinfo {author}
  {\bibfnamefont {R.}~\bibnamefont {Li}}, \bibinfo {author} {\bibfnamefont
  {Y.}~\bibnamefont {Wang}}, \bibinfo {author} {\bibfnamefont {Q.}~\bibnamefont
  {Zhang}}, \bibinfo {author} {\bibfnamefont {F.}~\bibnamefont {Meng}},
  \bibinfo {author} {\bibfnamefont {L.}~\bibnamefont {Gu}}, \bibinfo {author}
  {\bibfnamefont {K.}~\bibnamefont {Zhu}}, \bibinfo {author} {\bibfnamefont
  {Y.}~\bibnamefont {Zang}}, \bibinfo {author} {\bibfnamefont {L.}~\bibnamefont
  {Li}}, \bibinfo {author} {\bibfnamefont {R.}~\bibnamefont {Shang}}, \bibinfo
  {author} {\bibfnamefont {X.}~\bibnamefont {Feng}}, \bibinfo {author}
  {\bibfnamefont {Q.-K.}\ \bibnamefont {Xue}}, \bibinfo {author} {\bibfnamefont
  {K.}~\bibnamefont {He}},\ and\ \bibinfo {author} {\bibfnamefont
  {H.}~\bibnamefont {Zhang}},\ }\bibfield  {title} {\bibinfo {title}
  {Observation of {A}haronov-{B}ohm effect in {P}b{T}e nanowire networks},\
  }\href {https://doi.org/10.1103/PhysRevB.105.L241112} {\bibfield  {journal}
  {\bibinfo  {journal} {Phys. Rev. B}\ }\textbf {\bibinfo {volume} {105}},\
  \bibinfo {pages} {L241112} (\bibinfo {year} {2022})}\BibitemShut {NoStop}%
\bibitem [{\citenamefont {Zhang}\ \emph {et~al.}(2023)\citenamefont {Zhang},
  \citenamefont {Song}, \citenamefont {Gao}, \citenamefont {Wang},
  \citenamefont {Yu}, \citenamefont {Yang}, \citenamefont {Jiang},
  \citenamefont {Miao}, \citenamefont {Li}, \citenamefont {Chen}, \citenamefont
  {Geng}, \citenamefont {Zhang}, \citenamefont {Meng}, \citenamefont {Lin},
  \citenamefont {Gu}, \citenamefont {Zhu}, \citenamefont {Zang}, \citenamefont
  {Li}, \citenamefont {Shang}, \citenamefont {Feng}, \citenamefont {Xue},
  \citenamefont {He},\ and\ \citenamefont {Zhang}}]{Zitong}%
  \BibitemOpen
  \bibfield  {author} {\bibinfo {author} {\bibfnamefont {Z.}~\bibnamefont
  {Zhang}}, \bibinfo {author} {\bibfnamefont {W.}~\bibnamefont {Song}},
  \bibinfo {author} {\bibfnamefont {Y.}~\bibnamefont {Gao}}, \bibinfo {author}
  {\bibfnamefont {Y.}~\bibnamefont {Wang}}, \bibinfo {author} {\bibfnamefont
  {Z.}~\bibnamefont {Yu}}, \bibinfo {author} {\bibfnamefont {S.}~\bibnamefont
  {Yang}}, \bibinfo {author} {\bibfnamefont {Y.}~\bibnamefont {Jiang}},
  \bibinfo {author} {\bibfnamefont {W.}~\bibnamefont {Miao}}, \bibinfo {author}
  {\bibfnamefont {R.}~\bibnamefont {Li}}, \bibinfo {author} {\bibfnamefont
  {F.}~\bibnamefont {Chen}}, \bibinfo {author} {\bibfnamefont {Z.}~\bibnamefont
  {Geng}}, \bibinfo {author} {\bibfnamefont {Q.}~\bibnamefont {Zhang}},
  \bibinfo {author} {\bibfnamefont {F.}~\bibnamefont {Meng}}, \bibinfo {author}
  {\bibfnamefont {T.}~\bibnamefont {Lin}}, \bibinfo {author} {\bibfnamefont
  {L.}~\bibnamefont {Gu}}, \bibinfo {author} {\bibfnamefont {K.}~\bibnamefont
  {Zhu}}, \bibinfo {author} {\bibfnamefont {Y.}~\bibnamefont {Zang}}, \bibinfo
  {author} {\bibfnamefont {L.}~\bibnamefont {Li}}, \bibinfo {author}
  {\bibfnamefont {R.}~\bibnamefont {Shang}}, \bibinfo {author} {\bibfnamefont
  {X.}~\bibnamefont {Feng}}, \bibinfo {author} {\bibfnamefont {Q.-K.}\
  \bibnamefont {Xue}}, \bibinfo {author} {\bibfnamefont {K.}~\bibnamefont
  {He}},\ and\ \bibinfo {author} {\bibfnamefont {H.}~\bibnamefont {Zhang}},\
  }\bibfield  {title} {\bibinfo {title} {Proximity effect in {PbTe-Pb} hybrid
  nanowire {J}osephson junctions},\ }\href
  {https://doi.org/10.1103/PhysRevMaterials.7.086201} {\bibfield  {journal}
  {\bibinfo  {journal} {Phys. Rev. Mater.}\ }\textbf {\bibinfo {volume} {7}},\
  \bibinfo {pages} {086201} (\bibinfo {year} {2023})}\BibitemShut {NoStop}%
\bibitem [{\citenamefont {Gao}\ \emph {et~al.}(2024{\natexlab{a}})\citenamefont
  {Gao}, \citenamefont {Song}, \citenamefont {Yang}, \citenamefont {Yu},
  \citenamefont {Li}, \citenamefont {Miao}, \citenamefont {Wang}, \citenamefont
  {Chen}, \citenamefont {Geng}, \citenamefont {Yang}, \citenamefont {Xia},
  \citenamefont {Feng}, \citenamefont {Zang}, \citenamefont {Li}, \citenamefont
  {Shang}, \citenamefont {Xue}, \citenamefont {He},\ and\ \citenamefont
  {Zhang}}]{Yichun_Gap}%
  \BibitemOpen
  \bibfield  {author} {\bibinfo {author} {\bibfnamefont {Y.}~\bibnamefont
  {Gao}}, \bibinfo {author} {\bibfnamefont {W.}~\bibnamefont {Song}}, \bibinfo
  {author} {\bibfnamefont {S.}~\bibnamefont {Yang}}, \bibinfo {author}
  {\bibfnamefont {Z.}~\bibnamefont {Yu}}, \bibinfo {author} {\bibfnamefont
  {R.}~\bibnamefont {Li}}, \bibinfo {author} {\bibfnamefont {W.}~\bibnamefont
  {Miao}}, \bibinfo {author} {\bibfnamefont {Y.}~\bibnamefont {Wang}}, \bibinfo
  {author} {\bibfnamefont {F.}~\bibnamefont {Chen}}, \bibinfo {author}
  {\bibfnamefont {Z.}~\bibnamefont {Geng}}, \bibinfo {author} {\bibfnamefont
  {L.}~\bibnamefont {Yang}}, \bibinfo {author} {\bibfnamefont {Z.}~\bibnamefont
  {Xia}}, \bibinfo {author} {\bibfnamefont {X.}~\bibnamefont {Feng}}, \bibinfo
  {author} {\bibfnamefont {Y.}~\bibnamefont {Zang}}, \bibinfo {author}
  {\bibfnamefont {L.}~\bibnamefont {Li}}, \bibinfo {author} {\bibfnamefont
  {R.}~\bibnamefont {Shang}}, \bibinfo {author} {\bibfnamefont {Q.-K.}\
  \bibnamefont {Xue}}, \bibinfo {author} {\bibfnamefont {K.}~\bibnamefont
  {He}},\ and\ \bibinfo {author} {\bibfnamefont {H.}~\bibnamefont {Zhang}},\
  }\bibfield  {title} {\bibinfo {title} {Hard superconducting gap in {PbTe}
  nanowires},\ }\href {https://doi.org/10.1088/0256-307X/41/3/038502}
  {\bibfield  {journal} {\bibinfo  {journal} {Chinese Physics Letters}\
  }\textbf {\bibinfo {volume} {41}},\ \bibinfo {pages} {038502} (\bibinfo
  {year} {2024}{\natexlab{a}})}\BibitemShut {NoStop}%
\bibitem [{\citenamefont {Li}\ \emph {et~al.}(2024)\citenamefont {Li},
  \citenamefont {Song}, \citenamefont {Miao}, \citenamefont {Yu}, \citenamefont
  {Wang}, \citenamefont {Yang}, \citenamefont {Gao}, \citenamefont {Wang},
  \citenamefont {Chen}, \citenamefont {Geng}, \citenamefont {Yang},
  \citenamefont {Xu}, \citenamefont {Feng}, \citenamefont {Wang}, \citenamefont
  {Zang}, \citenamefont {Li}, \citenamefont {Shang}, \citenamefont {Xue},
  \citenamefont {He},\ and\ \citenamefont {Zhang}}]{Ruidong_Planar}%
  \BibitemOpen
  \bibfield  {author} {\bibinfo {author} {\bibfnamefont {R.}~\bibnamefont
  {Li}}, \bibinfo {author} {\bibfnamefont {W.}~\bibnamefont {Song}}, \bibinfo
  {author} {\bibfnamefont {W.}~\bibnamefont {Miao}}, \bibinfo {author}
  {\bibfnamefont {Z.}~\bibnamefont {Yu}}, \bibinfo {author} {\bibfnamefont
  {Z.}~\bibnamefont {Wang}}, \bibinfo {author} {\bibfnamefont {S.}~\bibnamefont
  {Yang}}, \bibinfo {author} {\bibfnamefont {Y.}~\bibnamefont {Gao}}, \bibinfo
  {author} {\bibfnamefont {Y.}~\bibnamefont {Wang}}, \bibinfo {author}
  {\bibfnamefont {F.}~\bibnamefont {Chen}}, \bibinfo {author} {\bibfnamefont
  {Z.}~\bibnamefont {Geng}}, \bibinfo {author} {\bibfnamefont {L.}~\bibnamefont
  {Yang}}, \bibinfo {author} {\bibfnamefont {J.}~\bibnamefont {Xu}}, \bibinfo
  {author} {\bibfnamefont {X.}~\bibnamefont {Feng}}, \bibinfo {author}
  {\bibfnamefont {T.}~\bibnamefont {Wang}}, \bibinfo {author} {\bibfnamefont
  {Y.}~\bibnamefont {Zang}}, \bibinfo {author} {\bibfnamefont {L.}~\bibnamefont
  {Li}}, \bibinfo {author} {\bibfnamefont {R.}~\bibnamefont {Shang}}, \bibinfo
  {author} {\bibfnamefont {Q.}~\bibnamefont {Xue}}, \bibinfo {author}
  {\bibfnamefont {K.}~\bibnamefont {He}},\ and\ \bibinfo {author}
  {\bibfnamefont {H.}~\bibnamefont {Zhang}},\ }\bibfield  {title} {\bibinfo
  {title} {Selective-area-grown {PbTe-Pb} planar {J}osephson junctions for
  quantum devices},\ }\href {https://doi.org/10.1021/acs.nanolett.4c00900}
  {\bibfield  {journal} {\bibinfo  {journal} {Nano Letters}\ }\textbf {\bibinfo
  {volume} {24}},\ \bibinfo {pages} {4658} (\bibinfo {year}
  {2024})}\BibitemShut {NoStop}%
\bibitem [{\citenamefont {Gupta}\ \emph {et~al.}(2024)\citenamefont {Gupta},
  \citenamefont {Khade}, \citenamefont {Riggert}, \citenamefont {Shani},
  \citenamefont {Menning}, \citenamefont {Lueb}, \citenamefont {Jung},
  \citenamefont {M{\'e}lin}, \citenamefont {Bakkers},\ and\ \citenamefont
  {Pribiag}}]{Vlad_PbTe}%
  \BibitemOpen
  \bibfield  {author} {\bibinfo {author} {\bibfnamefont {M.}~\bibnamefont
  {Gupta}}, \bibinfo {author} {\bibfnamefont {V.}~\bibnamefont {Khade}},
  \bibinfo {author} {\bibfnamefont {C.}~\bibnamefont {Riggert}}, \bibinfo
  {author} {\bibfnamefont {L.}~\bibnamefont {Shani}}, \bibinfo {author}
  {\bibfnamefont {G.}~\bibnamefont {Menning}}, \bibinfo {author} {\bibfnamefont
  {P.~J.~H.}\ \bibnamefont {Lueb}}, \bibinfo {author} {\bibfnamefont
  {J.}~\bibnamefont {Jung}}, \bibinfo {author} {\bibfnamefont {R.}~\bibnamefont
  {M{\'e}lin}}, \bibinfo {author} {\bibfnamefont {E.~P. A.~M.}\ \bibnamefont
  {Bakkers}},\ and\ \bibinfo {author} {\bibfnamefont {V.~S.}\ \bibnamefont
  {Pribiag}},\ }\bibfield  {title} {\bibinfo {title} {Evidence for
  $\pi$-shifted cooper quartets and few-mode transport in pbte nanowire
  three-terminal josephson junctions},\ }\href
  {https://doi.org/10.1021/acs.nanolett.4c02414} {\bibfield  {journal}
  {\bibinfo  {journal} {Nano Letters}\ }\textbf {\bibinfo {volume} {24}},\
  \bibinfo {pages} {13903} (\bibinfo {year} {2024})}\BibitemShut {NoStop}%
\bibitem [{\citenamefont {Geng}\ \emph {et~al.}(2025)\citenamefont {Geng},
  \citenamefont {Chen}, \citenamefont {Gao}, \citenamefont {Yang},
  \citenamefont {Wang}, \citenamefont {Yang}, \citenamefont {Zhang},
  \citenamefont {Li}, \citenamefont {Song}, \citenamefont {Xu}, \citenamefont
  {Yu}, \citenamefont {Li}, \citenamefont {Wang}, \citenamefont {Feng},
  \citenamefont {Wang}, \citenamefont {Zang}, \citenamefont {Li}, \citenamefont
  {Shang}, \citenamefont {Xue}, \citenamefont {He},\ and\ \citenamefont
  {Zhang}}]{PbTe_In}%
  \BibitemOpen
  \bibfield  {author} {\bibinfo {author} {\bibfnamefont {Z.}~\bibnamefont
  {Geng}}, \bibinfo {author} {\bibfnamefont {F.}~\bibnamefont {Chen}}, \bibinfo
  {author} {\bibfnamefont {Y.}~\bibnamefont {Gao}}, \bibinfo {author}
  {\bibfnamefont {L.}~\bibnamefont {Yang}}, \bibinfo {author} {\bibfnamefont
  {Y.}~\bibnamefont {Wang}}, \bibinfo {author} {\bibfnamefont {S.}~\bibnamefont
  {Yang}}, \bibinfo {author} {\bibfnamefont {S.}~\bibnamefont {Zhang}},
  \bibinfo {author} {\bibfnamefont {Z.}~\bibnamefont {Li}}, \bibinfo {author}
  {\bibfnamefont {W.}~\bibnamefont {Song}}, \bibinfo {author} {\bibfnamefont
  {J.}~\bibnamefont {Xu}}, \bibinfo {author} {\bibfnamefont {Z.}~\bibnamefont
  {Yu}}, \bibinfo {author} {\bibfnamefont {R.}~\bibnamefont {Li}}, \bibinfo
  {author} {\bibfnamefont {Z.}~\bibnamefont {Wang}}, \bibinfo {author}
  {\bibfnamefont {X.}~\bibnamefont {Feng}}, \bibinfo {author} {\bibfnamefont
  {T.}~\bibnamefont {Wang}}, \bibinfo {author} {\bibfnamefont {Y.}~\bibnamefont
  {Zang}}, \bibinfo {author} {\bibfnamefont {L.}~\bibnamefont {Li}}, \bibinfo
  {author} {\bibfnamefont {R.}~\bibnamefont {Shang}}, \bibinfo {author}
  {\bibfnamefont {Q.-K.}\ \bibnamefont {Xue}}, \bibinfo {author} {\bibfnamefont
  {K.}~\bibnamefont {He}},\ and\ \bibinfo {author} {\bibfnamefont
  {H.}~\bibnamefont {Zhang}},\ }\bibfield  {title} {\bibinfo {title} {Enhanced
  superconductivity in {PbTe-In} hybrids},\ }\href
  {https://doi.org/10.1103/8czc-tn4z} {\bibfield  {journal} {\bibinfo
  {journal} {Phys. Rev. Mater.}\ }\textbf {\bibinfo {volume} {9}},\ \bibinfo
  {pages} {084802} (\bibinfo {year} {2025})}\BibitemShut {NoStop}%
\bibitem [{\citenamefont {Gao}\ \emph {et~al.}(2024{\natexlab{b}})\citenamefont
  {Gao}, \citenamefont {Song}, \citenamefont {Yu}, \citenamefont {Yang},
  \citenamefont {Wang}, \citenamefont {Li}, \citenamefont {Chen}, \citenamefont
  {Geng}, \citenamefont {Yang}, \citenamefont {Xu} \emph
  {et~al.}}]{Yichun_SQUID}%
  \BibitemOpen
  \bibfield  {author} {\bibinfo {author} {\bibfnamefont {Y.}~\bibnamefont
  {Gao}}, \bibinfo {author} {\bibfnamefont {W.}~\bibnamefont {Song}}, \bibinfo
  {author} {\bibfnamefont {Z.}~\bibnamefont {Yu}}, \bibinfo {author}
  {\bibfnamefont {S.}~\bibnamefont {Yang}}, \bibinfo {author} {\bibfnamefont
  {Y.}~\bibnamefont {Wang}}, \bibinfo {author} {\bibfnamefont {R.}~\bibnamefont
  {Li}}, \bibinfo {author} {\bibfnamefont {F.}~\bibnamefont {Chen}}, \bibinfo
  {author} {\bibfnamefont {Z.}~\bibnamefont {Geng}}, \bibinfo {author}
  {\bibfnamefont {L.}~\bibnamefont {Yang}}, \bibinfo {author} {\bibfnamefont
  {J.}~\bibnamefont {Xu}}, \emph {et~al.},\ }\bibfield  {title} {\bibinfo
  {title} {{SQUID} oscillations in {PbTe} nanowire networks},\ }\href
  {https://doi.org/10.1103/PhysRevB.110.045405} {\bibfield  {journal} {\bibinfo
   {journal} {Phys. Rev. B}\ }\textbf {\bibinfo {volume} {110}},\ \bibinfo
  {pages} {045405} (\bibinfo {year} {2024}{\natexlab{b}})}\BibitemShut
  {NoStop}%
\bibitem [{SM()}]{SM}%
  \BibitemOpen
  \href@noop {} {\bibinfo {title} {See {S}upplemental {M}aterial at [url link]
  for additional data and analysis.}}\BibitemShut {Stop}%
\bibitem [{\citenamefont {Grabecki}\ \emph {et~al.}(2004)\citenamefont
  {Grabecki}, \citenamefont {Wróbel}, \citenamefont {Dietl}, \citenamefont
  {Polakowska}, \citenamefont {Kaminska}, \citenamefont {Piotrowska},
  \citenamefont {Ratuszna}, \citenamefont {Springholz},\ and\ \citenamefont
  {Bauer}}]{Physica_E_2004_PbTe_QPC}%
  \BibitemOpen
  \bibfield  {author} {\bibinfo {author} {\bibfnamefont {G.}~\bibnamefont
  {Grabecki}}, \bibinfo {author} {\bibfnamefont {J.}~\bibnamefont {Wróbel}},
  \bibinfo {author} {\bibfnamefont {T.}~\bibnamefont {Dietl}}, \bibinfo
  {author} {\bibfnamefont {E.}~\bibnamefont {Polakowska}}, \bibinfo {author}
  {\bibfnamefont {E.}~\bibnamefont {Kaminska}}, \bibinfo {author}
  {\bibfnamefont {A.}~\bibnamefont {Piotrowska}}, \bibinfo {author}
  {\bibfnamefont {A.}~\bibnamefont {Ratuszna}}, \bibinfo {author}
  {\bibfnamefont {G.}~\bibnamefont {Springholz}},\ and\ \bibinfo {author}
  {\bibfnamefont {G.}~\bibnamefont {Bauer}},\ }\bibfield  {title} {\bibinfo
  {title} {Ballistic transport in {P}b{T}e-based nanostructures},\ }\href
  {https://doi.org/10.1016/j.physe.2003.08.010} {\bibfield  {journal} {\bibinfo
   {journal} {Physica E: Low-dimensional Systems and Nanostructures}\ }\textbf
  {\bibinfo {volume} {20}},\ \bibinfo {pages} {236} (\bibinfo {year}
  {2004})}\BibitemShut {NoStop}%
\bibitem [{\citenamefont {Antipov}\ \emph {et~al.}(2018)\citenamefont
  {Antipov}, \citenamefont {Bargerbos}, \citenamefont {Winkler}, \citenamefont
  {Bauer}, \citenamefont {Rossi},\ and\ \citenamefont
  {Lutchyn}}]{LutchynSchrodinger}%
  \BibitemOpen
  \bibfield  {author} {\bibinfo {author} {\bibfnamefont {A.~E.}\ \bibnamefont
  {Antipov}}, \bibinfo {author} {\bibfnamefont {A.}~\bibnamefont {Bargerbos}},
  \bibinfo {author} {\bibfnamefont {G.~W.}\ \bibnamefont {Winkler}}, \bibinfo
  {author} {\bibfnamefont {B.}~\bibnamefont {Bauer}}, \bibinfo {author}
  {\bibfnamefont {E.}~\bibnamefont {Rossi}},\ and\ \bibinfo {author}
  {\bibfnamefont {R.~M.}\ \bibnamefont {Lutchyn}},\ }\bibfield  {title}
  {\bibinfo {title} {Effects of gate-induced electric fields on semiconductor
  {M}ajorana nanowires},\ }\href {https://doi.org/10.1103/PhysRevX.8.031041}
  {\bibfield  {journal} {\bibinfo  {journal} {Phys. Rev. X}\ }\textbf {\bibinfo
  {volume} {8}},\ \bibinfo {pages} {031041} (\bibinfo {year}
  {2018})}\BibitemShut {NoStop}%
\bibitem [{\citenamefont {Mikkelsen}\ \emph {et~al.}(2018)\citenamefont
  {Mikkelsen}, \citenamefont {Kotetes}, \citenamefont {Krogstrup},\ and\
  \citenamefont {Flensberg}}]{Flensberg_SP}%
  \BibitemOpen
  \bibfield  {author} {\bibinfo {author} {\bibfnamefont {A.~E.~G.}\
  \bibnamefont {Mikkelsen}}, \bibinfo {author} {\bibfnamefont {P.}~\bibnamefont
  {Kotetes}}, \bibinfo {author} {\bibfnamefont {P.}~\bibnamefont {Krogstrup}},\
  and\ \bibinfo {author} {\bibfnamefont {K.}~\bibnamefont {Flensberg}},\
  }\bibfield  {title} {\bibinfo {title} {Hybridization at
  superconductor-semiconductor interfaces},\ }\href
  {https://doi.org/10.1103/PhysRevX.8.031040} {\bibfield  {journal} {\bibinfo
  {journal} {Phys. Rev. X}\ }\textbf {\bibinfo {volume} {8}},\ \bibinfo {pages}
  {031040} (\bibinfo {year} {2018})}\BibitemShut {NoStop}%
\bibitem [{\citenamefont {Chen}\ \emph {et~al.}(2023)\citenamefont {Chen},
  \citenamefont {Cao}, \citenamefont {He}, \citenamefont {Liu},\ and\
  \citenamefont {Liu}}]{ChenLi_Simulation}%
  \BibitemOpen
  \bibfield  {author} {\bibinfo {author} {\bibfnamefont {L.}~\bibnamefont
  {Chen}}, \bibinfo {author} {\bibfnamefont {Z.}~\bibnamefont {Cao}}, \bibinfo
  {author} {\bibfnamefont {K.}~\bibnamefont {He}}, \bibinfo {author}
  {\bibfnamefont {X.}~\bibnamefont {Liu}},\ and\ \bibinfo {author}
  {\bibfnamefont {D.~E.}\ \bibnamefont {Liu}},\ }\bibfield  {title} {\bibinfo
  {title} {Electrostatic effects of
  {${\mathrm{{MnBi}}}_{2}{\mathrm{{Te}}}_{4}$}-superconductor heterostructures
  in the chiral majorana search},\ }\href
  {https://doi.org/10.1103/PhysRevB.107.165405} {\bibfield  {journal} {\bibinfo
   {journal} {Phys. Rev. B}\ }\textbf {\bibinfo {volume} {107}},\ \bibinfo
  {pages} {165405} (\bibinfo {year} {2023})}\BibitemShut {NoStop}%
\bibitem [{\citenamefont {de~Andrada~e Silva}(1999)}]{PbTe_parameter1}%
  \BibitemOpen
  \bibfield  {author} {\bibinfo {author} {\bibfnamefont {E.~A.}\ \bibnamefont
  {de~Andrada~e Silva}},\ }\bibfield  {title} {\bibinfo {title} {Optical
  transition energies for lead-salt semiconductor quantum wells},\ }\href
  {https://doi.org/10.1103/PhysRevB.60.8859} {\bibfield  {journal} {\bibinfo
  {journal} {Phys. Rev. B}\ }\textbf {\bibinfo {volume} {60}},\ \bibinfo
  {pages} {8859} (\bibinfo {year} {1999})}\BibitemShut {NoStop}%
\bibitem [{\citenamefont {Ridolfi}\ \emph {et~al.}(2015)\citenamefont
  {Ridolfi}, \citenamefont {Silva},\ and\ \citenamefont
  {La~Rocca}}]{PbTe_parameter2}%
  \BibitemOpen
  \bibfield  {author} {\bibinfo {author} {\bibfnamefont {E.}~\bibnamefont
  {Ridolfi}}, \bibinfo {author} {\bibfnamefont {E.~A. d. A.~e.}\ \bibnamefont
  {Silva}},\ and\ \bibinfo {author} {\bibfnamefont {G.~C.}\ \bibnamefont
  {La~Rocca}},\ }\bibfield  {title} {\bibinfo {title} {Effective $g$-factor
  tensor for carriers in {IV-VI} semiconductor quantum wells},\ }\href
  {https://doi.org/10.1103/PhysRevB.91.085313} {\bibfield  {journal} {\bibinfo
  {journal} {Phys. Rev. B}\ }\textbf {\bibinfo {volume} {91}},\ \bibinfo
  {pages} {085313} (\bibinfo {year} {2015})}\BibitemShut {NoStop}%
\bibitem [{\citenamefont {Zhang}\ \emph {et~al.}(2019)\citenamefont {Zhang},
  \citenamefont {Liu}, \citenamefont {Wimmer},\ and\ \citenamefont
  {Kouwenhoven}}]{NextSteps}%
  \BibitemOpen
  \bibfield  {author} {\bibinfo {author} {\bibfnamefont {H.}~\bibnamefont
  {Zhang}}, \bibinfo {author} {\bibfnamefont {D.~E.}\ \bibnamefont {Liu}},
  \bibinfo {author} {\bibfnamefont {M.}~\bibnamefont {Wimmer}},\ and\ \bibinfo
  {author} {\bibfnamefont {L.~P.}\ \bibnamefont {Kouwenhoven}},\ }\bibfield
  {title} {\bibinfo {title} {Next steps of quantum transport in {M}ajorana
  nanowire devices},\ }\href {https://doi.org/10.1038/s41467-019-13133-1}
  {\bibfield  {journal} {\bibinfo  {journal} {Nature Communications}\ }\textbf
  {\bibinfo {volume} {10}},\ \bibinfo {pages} {5128} (\bibinfo {year}
  {2019})}\BibitemShut {NoStop}%
\bibitem [{\citenamefont {Prada}\ \emph {et~al.}(2020)\citenamefont {Prada},
  \citenamefont {San-Jose}, \citenamefont {de~Moor}, \citenamefont {Geresdi},
  \citenamefont {Lee}, \citenamefont {Klinovaja}, \citenamefont {Loss},
  \citenamefont {Nyg{\aa}rd}, \citenamefont {Aguado},\ and\ \citenamefont
  {Kouwenhoven}}]{Prada2020}%
  \BibitemOpen
  \bibfield  {author} {\bibinfo {author} {\bibfnamefont {E.}~\bibnamefont
  {Prada}}, \bibinfo {author} {\bibfnamefont {P.}~\bibnamefont {San-Jose}},
  \bibinfo {author} {\bibfnamefont {M.~W.}\ \bibnamefont {de~Moor}}, \bibinfo
  {author} {\bibfnamefont {A.}~\bibnamefont {Geresdi}}, \bibinfo {author}
  {\bibfnamefont {E.~J.}\ \bibnamefont {Lee}}, \bibinfo {author} {\bibfnamefont
  {J.}~\bibnamefont {Klinovaja}}, \bibinfo {author} {\bibfnamefont
  {D.}~\bibnamefont {Loss}}, \bibinfo {author} {\bibfnamefont {J.}~\bibnamefont
  {Nyg{\aa}rd}}, \bibinfo {author} {\bibfnamefont {R.}~\bibnamefont {Aguado}},\
  and\ \bibinfo {author} {\bibfnamefont {L.~P.}\ \bibnamefont {Kouwenhoven}},\
  }\bibfield  {title} {\bibinfo {title} {From {A}ndreev to {M}ajorana bound
  states in hybrid superconductor--semiconductor nanowires},\ }\href
  {https://doi.org/10.1038/s42254-020-0228-y} {\bibfield  {journal} {\bibinfo
  {journal} {Nature Reviews Physics}\ }\textbf {\bibinfo {volume} {2}},\
  \bibinfo {pages} {575} (\bibinfo {year} {2020})}\BibitemShut {NoStop}%
\bibitem [{\citenamefont {Kouwenhoven}(0)}]{Leo_perspective}%
  \BibitemOpen
  \bibfield  {author} {\bibinfo {author} {\bibfnamefont {L.}~\bibnamefont
  {Kouwenhoven}},\ }\bibfield  {title} {\bibinfo {title} {Perspective on
  {M}ajorana bound-states in hybrid superconductor-semiconductor nanowires},\
  }\href {https://doi.org/10.1142/S0217984925400020} {\bibfield  {journal}
  {\bibinfo  {journal} {Modern Physics Letters B}\ }\textbf {\bibinfo {volume}
  {0}},\ \bibinfo {pages} {2540002} (\bibinfo {year} {0})}\BibitemShut
  {NoStop}%
\bibitem [{\citenamefont {Wang}\ \emph {et~al.}(2026)\citenamefont {Wang} \emph
  {et~al.}}]{rawdata}%
  \BibitemOpen
  \bibfield  {author} {\bibinfo {author} {\bibfnamefont {Y.}~\bibnamefont
  {Wang}} \emph {et~al.},\ }\href@noop {} {\bibinfo {title} {Zenodo}},\
  \bibinfo {howpublished} {\url{https://doi.org/10.5281/zenodo.18629635}}
  (\bibinfo {year} {2026})\BibitemShut {NoStop}%
\end{thebibliography}%

\newpage

\onecolumngrid

\newpage


\section*{Supplementary material}

# Supplemental Materials for "Presence versus absence of charging energies in PbTe quantum dots"

Yuhao Wang,[1,2,*] Lining Yang,[1,*] Wenyu Song,[1,*] Li Chen,[3,*] Zehao Yu,[1] Xinchen He,[1] Zeyu Yan,[1] Jiaye Xu,[1] Ruidong Li,[1] Weizhao Wang,[1] Zonglin Li,[1] Shuai Yang,[1] Shan Zhang,[1] Xiao Feng,[1,4,5,6] Tiantian Wang,[4,6] Yunyi Zang,[4,6] Lin Li,[4] Runan Shang,[4,6] Qi-Kun Xue,[1,4,5,6,7] Ke He,[1,4,5,6,†] and Hao Zhang[2,‡]

[1]*State Key Laboratory of Low Dimensional Quantum Physics, Department of Physics, Tsinghua University, Beijing 100084, China*
[2]*College of Semiconductors, Southern University of Science and Technology, Shenzhen 518055, China*
[3]*School of Physics and Optoelectronic Engineering, Guangdong University of Technology, Guangzhou 510006, China*
[4]*Beijing Academy of Quantum Information Sciences, Beijing 100193, China*
[5]*Frontier Science Center for Quantum Information, Beijing 100084, China*
[6]*Hefei National Laboratory, Hefei 230088, China*
[7]*Department of Physics, Southern University of Science and Technology, Shenzhen 518055, China*

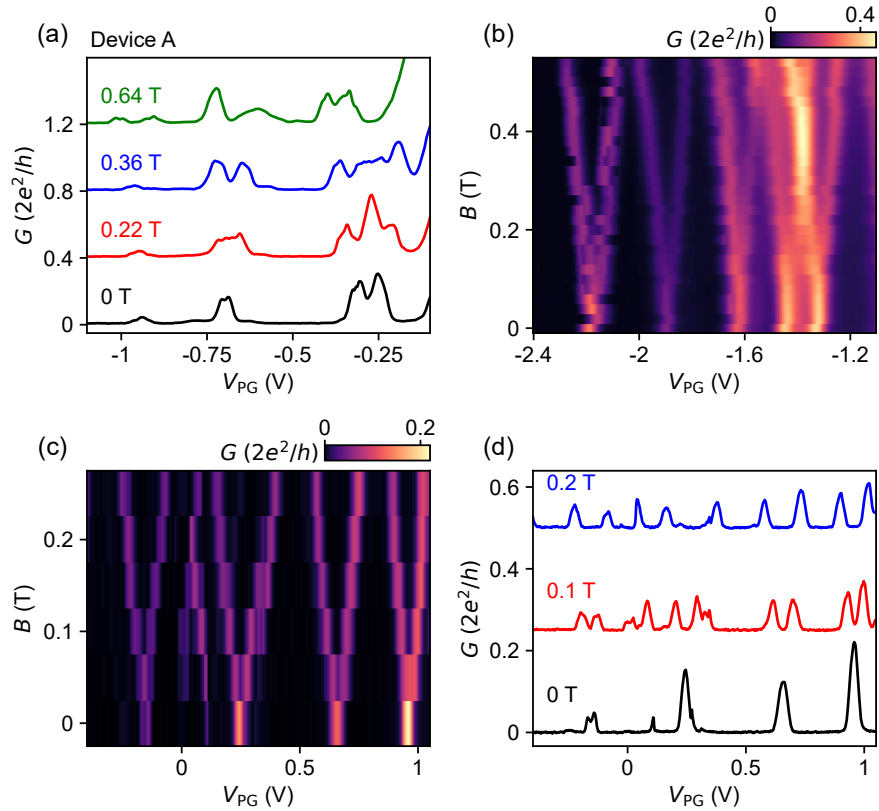

FIG. S1. Additional data for devices without $E_C$. (a) Line cuts extracted from Fig. 2(m) of device A. The double-peak feature near $V_{\text{PG}}$ = -0.7 V at 0 T does not indicate finite $E_C$ but arises from two nearly degenerate quantized levels. This double-peak feature splits into four peaks at 0.22 T, suggesting the absence of $E_C$. (b) The full $B$ scan data of Fig. 3(a). $V_{\text{TG1}}$ = -2 V, $V_{\text{TG2}}$ = -1 V. The $B$ direction is in-plane and perpendicular to the nanowire. (c) $B$ scan of another device with a cross-sectional area of 28300 nm$^2$. (d) Line cuts from (c). Note that the two devices in (b-d) have been studied in our previous work [1]. Here we show additional data from these two devices to investigate the relation between $E_C$ and cross-sectional area. The absence of $E_C$ in these two devices is consistent with their large cross-sectional areas.

---

[*] equal contribution
[†] kehe@tsinghua.edu.cn
[‡] zhanghao@sustech.edu.cn

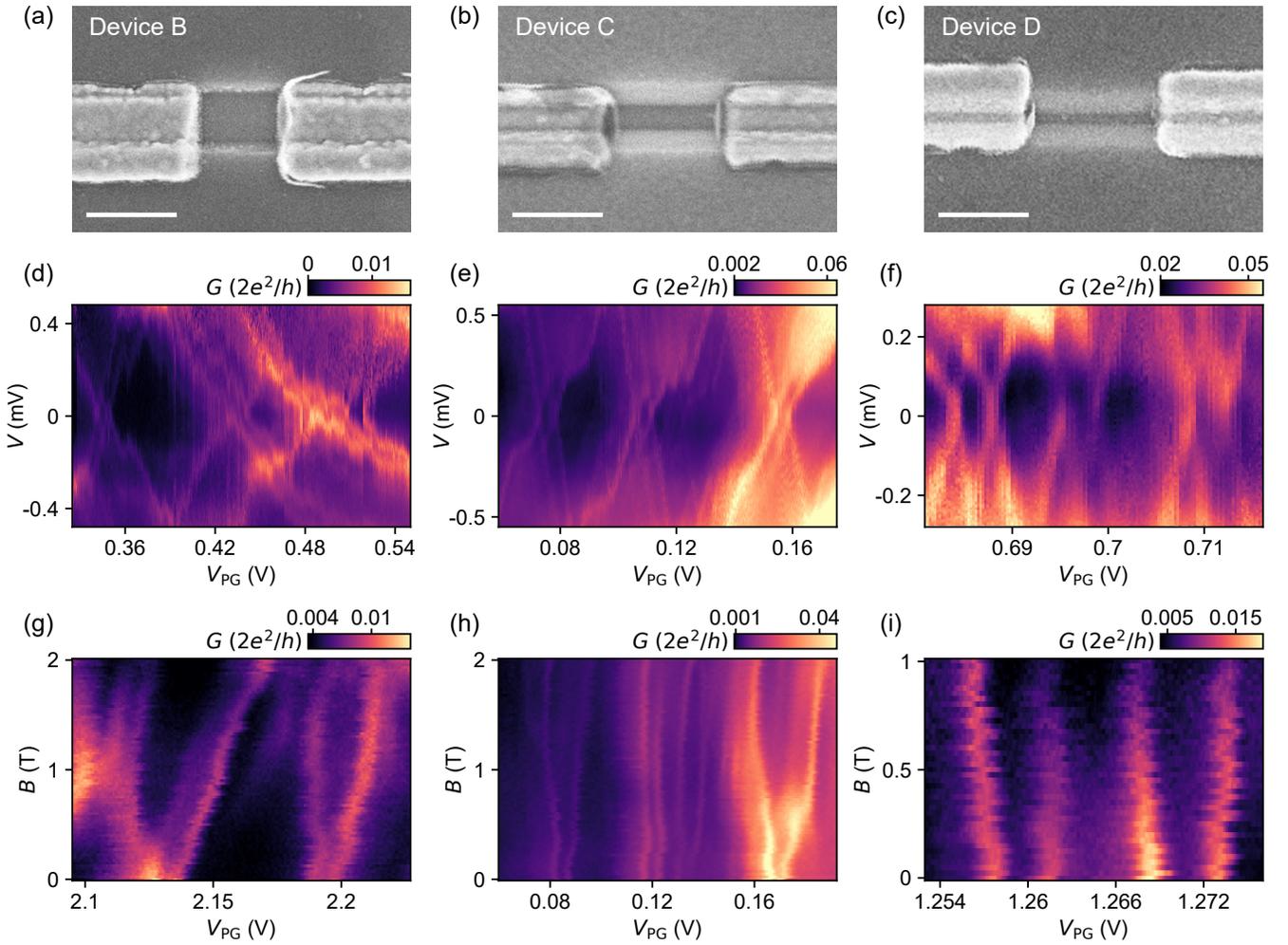

FIG. S2. Additional data for devices with $E_C$. (a-c) SEMs of devices B-D before the deposition of gates and dielectric layers. Scale bars are 300 nm. (d-f) Three additional charge stability diagrams of these three devices. The Coulomb diamonds alternate in size, suggesting the presence of $E_C$. (g-i) Three additional $B$ scans of these devices. In panel (g) (device B), the Coulomb blockade near $V_{PG} = 2.13$ V is not well defined. The tunnel gate settings, ($V_{TG1}$, $V_{TG2}$), for Figs. 2(j), 2(n), 2(k), 2(o), 2(l), 2(p) are (-0.5 V, -0.75 V), (-0.61 V, -1.025 V), (-3.56 V, -4.56 V), (-3.56 V, -4.56 V), (-0.63 V, -0.11 V), (-0.551 V, 0.601 V), respectively.

[1] Z. Li, W. Song, S. Zhang, Y. Wang, Z. Wang, Z. Yu, R. Li, Z. Yan, J. Xu, Y. Gao, S. Yang, L. Yang, X. Feng, T. Wang, Y. Zang, L. Li, R. Shang, Q.-K. Xue, K. He, and H. Zhang, Anisotropy of PbTe nanowires with and without a superconductor, Phys. Rev. B **111**, 195416 (2025).

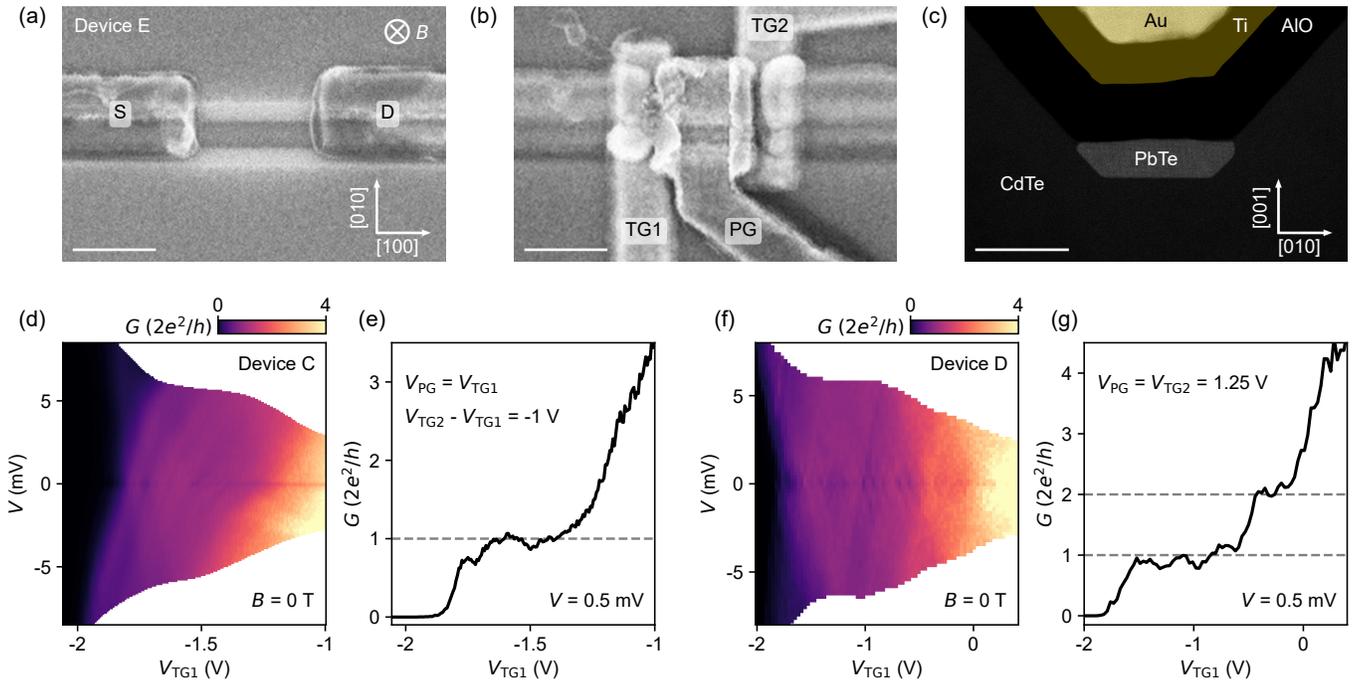

FIG. S3. Additional data for QPC-QD evolution. (a-b) SEMs of device E (Fig. 4) before (a) and after (b) the deposition of top gates. The scale bars are 300 nm. (c) STEM of device E. The scale bar is 50 nm. (d-e) Signature of zero-field ballistic transport in device C. The three gates were scanned simultaneously while keeping $V_{TG1} = V_{PG} = V_{TG2} + 1$ V. The $2e^2/h$ plateau is revealed as a violet diamond in (d). The near-zero-bias line cut ($V = 0.5$ mV) is shown in (e). (f-g) Ballistic signatures in device D. For devices C, D, and E, we have subtracted a contact resistance of 4.3 k$\Omega$, 3.6 k$\Omega$, and 3 k$\Omega$, respectively.

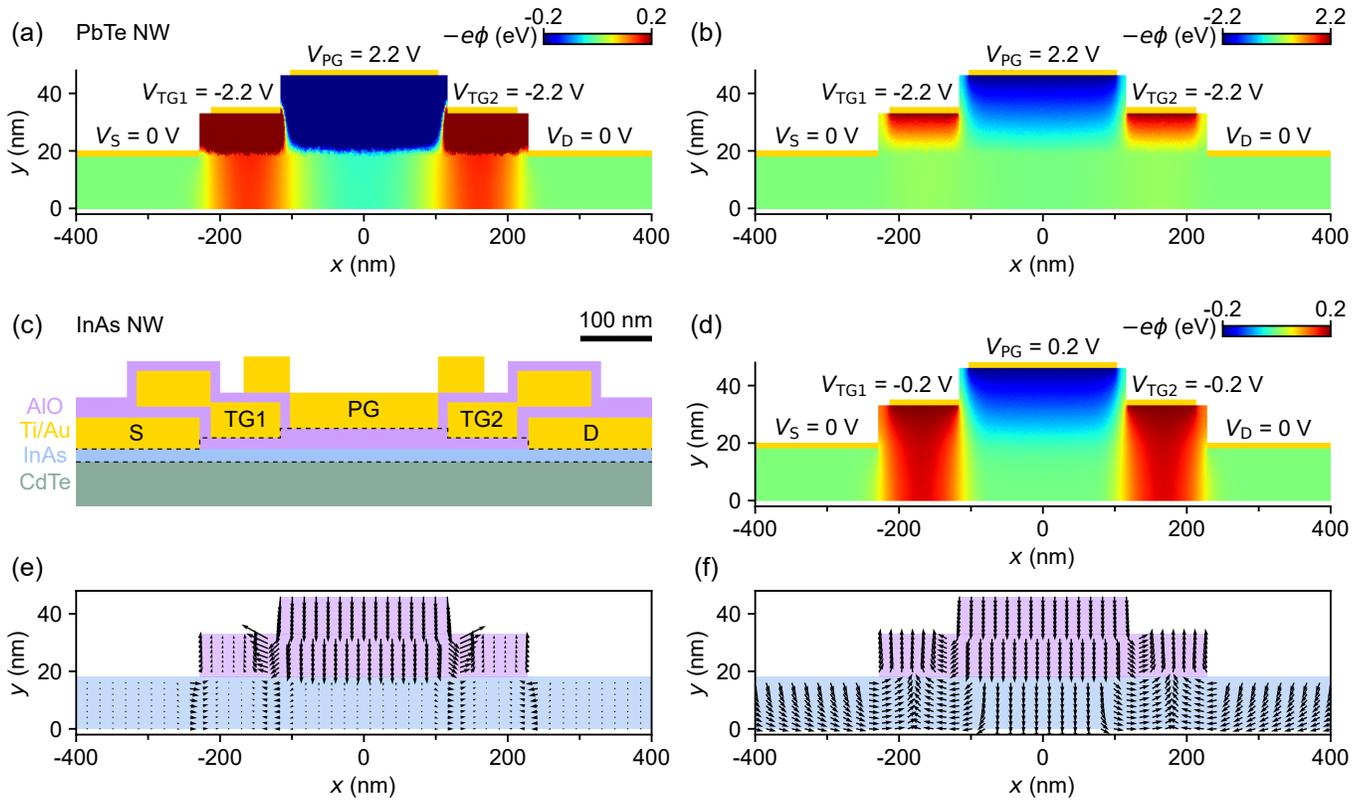

FIG. S4. Additional data for electrostatic simulations. (a-b) Replots of Fig. 5(b). The color ranges are (-0.2 eV, 0.2 eV) for (a) and (-2.2 eV, 2.2 eV) for (b), respectively. The former highlights the potential changes in the PbTe nanowire, while the latter shows the full potential variation across the device (including the dielectric layer). (c-f) Simulations for the InAs case (corresponding to the red curve in Fig. 5(f)). (c) The device geometry and dimension are identical to those of the PbTe case. (d) Simulated potential landscape for $V_{TG1} = V_{TG2} = -0.2$ V, $V_{PG} = 0.2$ V. The potential landscape in the nanowire region ($y < 18$ nm) is similar to that in (a). (e-f) Electric field distribution (e) and its normalized version (f).


\end{document}